\documentclass[a4paper]{jpconf}
\usepackage{amssymb}
\usepackage{amsmath}
\numberwithin{equation}{section}
\def\dd{{\rm d}}\def\ii{{\rm i}}
\def\SU{{\rm SU}}
\def\half{{\textstyle{\frac12}}}\def\third{{\textstyle{\frac13}}}
\def\fourth{{\textstyle{\frac14}}}\def\sixth{{\textstyle{\frac16}}}
\def\beq{\begin{equation}}\def\eeq{\end{equation}}
\def\bea{\begin{eqnarray}}\def\eea{\end{eqnarray}}

\begin{document}

\title{Dynamics for causal sets with matter fields:\\
A Lagrangian-based approach}

\author{Roman Sverdlov$^1$ and Luca Bombelli$^{2,3}$}
\address{$^1$ Physics Department, University of Michigan,
\\450 Church Street, Ann Arbor, MI 48109-1040, USA}
\address{$^2$ Department of Physics and Astronomy, University of Mississippi,
\\108 Lewis Hall, University, MS 38677-1848, USA}
\address{$^3$ Departament de F\'isica Fonamental, Universitat de Barcelona,
\\Av.\ Diagonal 647, 08028 Barcelona, Spain}
\ead{sverdlov$\_$roman@hotmail.com, bombelli@olemiss.edu}

\begin{abstract}
We present a framework for the dynamics of causal sets and coupled matter fields, which is a simplification and generalization of an approach we recently proposed. Given a set of fields including the gravitational one, the main step in implementing our proposal consists in writing their continuum-based action using as variables for the spacetime geometry the causal order and volume element. One then discretizes the resulting expression, with a procedure designed to maintain covariance. After a discussion of the general framework, we treat in detail the case of scalar fields, Yang-Mills gauge fields and the gravitational field.
\end{abstract}

\section{Introduction}

Despite the fact that the causal set program for quantum gravity \cite{CS} has been around for more than 20 years \cite{SorkinCausal, Dowker, Henson}, identifying an appropriate formulation of the dynamics of the theory remains an open issue. Such a formulation would allow us to recover a smooth Lorentzian geometry with matter fields in a suitable continuum approximation, and would have a semiclassical limit in which general relativity coupled to those matter fields, including possible quantum corrections, is recovered. Some proposals have been made, most notably the classical sequential growth dynamics models of Rideout and Sorkin \cite{SGD}, a class of stochastic growth models defined by the fact that they satisfy causal-set versions of the general covariance and causality requirements. While a number of interesting results have been obtained on these models, no way to recover the continuum geometry and dynamics from them is known, and no proposal has been made for how to include matter fields in them.

Here we will present a different approach to the dynamics of causal sets, one which is not based on general principles, but rather tries to mimick directly the continuum dynamics by using a discretized expression for the action, written in terms of variables that are meaningful for causal sets. Since a causal set is simply a locally finite partially ordered set, the only variables that characterize it are the {\em number\/} of elements in each region and the {\em order\/} in which they come. Thus, as far as gravity is concerned, our tasks are to express the Einstein-Hilbert action in terms of volumes of (finite) spacetime regions and causal relations, the continuum counterparts to the counting measure and the partial order, respectively, and then to discretize the result, making sure that we end up with a covariant expression.

But one of our main motivations for proposing this approach is that the action for other fields can be treated in the same way, as we will show below for scalar and gauge fields. The reason for including matter is not simply completeness, or even a desire to develop the phenomenology of the theory, but the belief that ultimately the gravitational field will have to be treated on the same footing as the other forms of matter from the beginning. Since all matter fields will be coupled to the geometry at the causal-set level, and most causal sets are not well approximated by smooth manifolds, it may be that not only the metric structure of spacetime, but even the possibility of defining topological and differentiable structures depends on the matter fields, that their presence is necessary in order to understand how manifoldlike causal sets emerge.

In section 2 we discuss the general framework and introduce the concept of Lagrangian generator, a building block from which suitable forms for the action of continuum fields can be written; section 3 treats the scalar field case and section 4 the case of a charged scalar field, i.e., one coupled to a gauge field; in section 5 we discuss the dynamics of the (Yang-Mills) gauge field itself, and in section 6 that of the gravitational field; the concluding section 7 contains some closing remarks. This article is based on the talk by LB at the DICE2008 conference, at which previous work along these lines \cite{paper1} was presented. After the talk was given, however, RS developed an improved version of this approach, and we felt that it would be more useful for this writeup to reflect the recent developments. For this reason, the material in this paper reflects to a large extent the content of one chapter of the PhD dissertation by RS.

\section{Lagrangians, pre-Lagrangians and Lagrangian generators}

It turns out that the causal-set versions of the Lagrangians all look somewhat complicated, but they are all complicated in the same way. To emphasize the pattern, we introduce the concept of Lagrangian generator, together with a prescription for going from a Lagrangian generator to the corresponding Lagrangian. Our aim in doing that is to ensure that Lagrangian generators for all known fields are simple-looking and the actual Lagrangians, as complicated as they might appear, can be ``read off" from the expression for the Lagrangian generator.

The key observation to guide us is the form in which one can write the scalars that appear in the usual continuum Lagrangian densities for the Klein-Gordon scalar, electromagnetic and gravitational fields. As will be shown in the next four sections, if $\alpha(p,q):= J^+(p) \cap J^-(q)$ is the Alexandrov set defined by two causally-related points $p \prec q$ in Minkowski space, then for any scalar fields and electromagnetic potentials varying linearly in inertial coordinates, respectively, the Lagrangian density at a point $x \in \alpha(p,q)$ can be written as
\bea
& &\kern-30pt k_{\rm sc}\, \partial^\mu\phi\, \partial_\mu\phi
= \frac1{\tau^2(p,q)}\, \big(\phi(q) - \phi(p)\big)^2
+ \frac{E_{\rm sc}}{\tau^{2d+2}(p,q)}\,
\int_{\alpha(p,q)} \dd^d r\,\dd^d s\, (\phi(r) - \phi(s))^2 \label{L1}\\
& &\kern-30pt k_{\rm em}\, F^{\mu\nu} F_{\mu\nu}
= \frac1{\tau^{d+2}(p,q)}\, \int_{\alpha(p,q)} \dd^d r\, \big(a(p,r)+a(r,q)+a(q,p)\big) + \nonumber\\
& &\kern39pt+\ \frac{E_{\rm em}}{\tau^{3d+2}(p,q)}\, \int_{\alpha(p,q)} \dd^d r\,\dd^d s\,\dd^d t\,
\big(a(r,s) + a(s,t) + a(t,r)\big)\;, \kern30pt \label{L2}
\eea
for appropriately adjusted pairs of coefficients $(k,E)$, while for any metric whose expansion in Riemann normal coordinates only contains quadratic non-trivial terms (the lowest possible order, so we will still refer informally to this situation as the ``linear case"),
\bea
& &\kern-16pt k_{\rm gr}\, R = \frac{V(\alpha(p,q))}{\tau^{d+2}(p,q)}
+ \frac{E_{\rm gr}}{\tau^{2d+2}(p,q)}\, \Big(\int_{\alpha(p,q)} \dd^d r\,
\big[V(\alpha(p,r)) + V(\alpha(r,q))\big] \Big)\,, \label{L3}
\eea
for appropriately adjusted values for $(k_{\rm gr},E_{\rm gr})$. Since $\alpha(p,q)$ is defined by causal relations alone without a direct reference to the continuum structure, and in particular observing that the right-hand sides of these integrals don't involve spacetime tensor indices, the replacement of integrals by sums makes it easy to generalize these expressions to a causal set.

A point that needs to be emphasized is that the expressions (\ref{L1})--(\ref{L3}) are covariantly defined scalars at a point, despite the fact that they involve the use of one Alexandrov set, as long as $E$ is chosen to have the correct, dimension-dependent value. This, however, brings up a problem when we take these expressions over to the causal-set context: in light of the fact that the theory is formulated for arbitrary causal sets and not just manifoldlike ones, if $E$ was viewed as a constant, it would be an extremely lucky coincidence that its value in the fundamental theory happened to be precisely the one needed for the exact cancellation of the non-covariant terms in the special case of a four-dimensional manifold. This, of course, is not satisfactory.

For this reason, we will view $E$ as a variable, subject to some physical laws that adjust its value to what it should be at any given point. The general idea is that the properties of a manifoldlike partial order determine the dimensionality of the manifold in which it can be faithfully embedded \cite{CS}, which in turn determines the value of $E$. More specifically, while in the linear manifoldlike scenario the only field $E$ depends on is gravity, as will be seen later in the non-linear case there is a slight dependence on the non-linear behavior of other fields, albeit very small. In the non-manifoldlike scenario, however, the dependence on non-gravitational fields might be arbitrarily large, just like the dependence on gravity would be. Consider a general theory for a set of fields of interest $F \in {\cal F}$, with a continuum Lagrangian density ${\cal L}(F,x,E)$ that can be obtained from expressions of the type (\ref{L1})--(\ref{L3}). Then
\beq
E = E(F,x)
\eeq
is a function both of the point $x$ and of the value of the fields $F \in {\cal F}$ in its neighborhood. Notice that since $E$ is not an independent field, a path integral for the field $F$ will not include a separate integration over $E$. Rather, it will be of the form
\beq
\int {\cal D}F \exp \Big(\ii \int \dd^d x\, {\cal L}\big(F,x,E(F,x)\big) \Big)\,.
\eeq

Let us now discuss how $E(F,x)$ is determined. The definition of $E$ for a general causal set requires a notion of degree of relativistic non-covariance for an expression such as (\ref{L1})--(\ref{L3}), which we will denote generically by ${\cal L}(F,E,p,q)$. If we identify a reference frame with the axis of an Alexandrov set (one defined by two points on the $t$-axis of the reference frame), then the degree of non-covariance represents the extent to which the value of ${\cal L}(F,E,p,q)$ depends on the choice of Alexandrov set $\alpha(p,q)$, and can be quantified as the range of variation of the values of ${\cal L}(F,E,p,q)$ as $p$ and $q$ are allowed to vary.

In the case of linear fields, the non-covariance is zero as long as $E$ is appropriately adjusted. In a curved Lorentzian geometry we can still have vanishing non-covariance, provided we go to the infinitesimal-$\alpha(p,q)$ limit. In the case of fields that can be seen as a discretized continuum it is not possible to select an infinitesimal region and the best one can do is constrain either the timelike length $\tau(p,q)$ or the volume $V(\alpha(p,q))$ between $p$ and $q$ to have a fixed (small) value, which means that some degree of non-covariance will appear, depending on the details of the discretization procedure; as long as the fields are linear or well-behaved functions that can be considered as slowly-varying, for an appropriately chosen $E$ the non-covariance should be small.

However, in the case of a general causal set, most of the assumptions that are made for the manifold can no longer be trusted, which means that it is possible for the degree of non-covariance to be large no matter what $E$ is selected to be. However, it is still possible to {\em formally\/} select $E$ in such a way that it minimizes the degree of non-covariance, even though the minimum might be very large.

One feature of Lorentzian geometries that complicates this procedure is that if the manifold is not compact the set of pairs $(p,q)$ may not be compact either, even when one imposes a condition on how far $p$ and $q$ are, so fields that are slowly varying in one Alexandrov set may appear to vary uncontrollably in another, highly boosted one that occupies a stretched region ``near the light cone of point $p$". Therefore, if all possible Alexandrov sets $\alpha(p,q)$ with $\tau(p,q) = \tau_0$ were considered, the value of $E$ might have nothing to do with the one it would have had in the linear case. For that reason, we constrain the varying Alexandrov set to vary within the boundaries of some other, larger ``fiducial" Alexandrov set $\alpha(P,Q)$, with
\beq
\tau_1 = \tau(p,q) < \tau(P,Q) = \tau_2\,.
\eeq
The variation, then, is defined by
\beq
{\rm var}_{\tau_1}({\cal L},F,E,P,Q):= \max \{ {\cal L}(F,E,p,q)
\mid P \prec p \prec q \prec Q\ \land\ \tau(p,q) = \tau_1 \}\,,
\eeq
and for any point $x$ in a causal set, $\alpha_{\tau_1,\tau_2}(F,x)$ is defined as the set of triples $(E,P,Q)$ for which ${\rm var}_{\tau_1}({\cal L},F,E,P,Q)$ is minimized with the constraint that $\tau(P,Q) = \tau_2$:
\bea
& &\kern-20pt\alpha_{\tau_1,\tau_2}(F,x) = \{(E,P,Q) \mid \tau(P,Q) = \tau_2\ \land\ \forall\ E',P',Q'\ {\rm with}\ \tau(P',Q') = \tau_2, \nonumber\\
& &\kern106pt {\rm var}_{\tau_1}({\cal L},F,E',P',Q') \geq {\rm var}_{\tau_1}({\cal L},F,E,P,Q)\}\,.
\eea
Typically, $\alpha_{\tau_1,\tau_2}(F,x)$ is a one-element set, and the Lagrangian at $x$ can simply be taken to be ${\cal L}(F,E,P,Q)$, where $(E,P,Q)$ is the unique element of that set. But in order to formally accommodate the cases where $\alpha_{\tau_1,\tau_2}(F,x)$ has more than one element, the Lagrangian at $x$ is formally defined as the average of the above over all elements of $\alpha_{\tau_1,\tau_2}(F,x)$:
\beq
{\cal L}_{\tau_1,\tau_2}(F,x) = \frac{1}{\sharp\{\alpha_{\tau_1,\tau_2}(F,x)\}}
\sum_{(E,P,Q) \in \alpha_{\tau_1,\tau_2}(F,x)} {\cal L}(F,E,P,Q)\,,
\eeq
where $\sharp\{\ \}$ denotes the cardinality of a set.

Now, dropping the issue of the value of $E$ and the selection of $p$ and $q$ for a second, let's switch gears and go back to the expression for ${\cal L}(F,E,p,q)$, which, from now on, will be referred to as a ``pre-Lagrangian". As the examples in the beginning of this section illustrate, the expressions for ${\cal L}(F,E,p,q)$ for scalar, gauge, and gravitational fields look similar, in the sense that they are all linear combinations of a function of $p$ and $q$ and some form of integral of the same function over the interior of $\alpha(p,q)$. So the natural question arises: why do Lagrangians take this particular form and not some other one? That question is answered by introducing the concept of Lagrangian generator and formally defining a procedure of going from a Lagrangian generator to a pre-Lagrangian in such a way that the above mentioned linear combination arises in a natural way if one formally follows the steps of the procedure. 

For simplicity, let's consider first an expression for a pre-Lagrangian of the general type (\ref{L3}),
\beq
{\cal L}_{\cal J}(F,E,p,q) = {\cal J}(p,q)
+ \frac{E}{\tau^d(p,q)} \int \dd^d r\, {\cal J}(p,r)\,, \label{LJ3}
\eeq
where $\cal J$ is a real-valued function depending on the set of fields $F$, which maps $(p,q)\mapsto {\cal J}(F,p,q) \in\mathbb{R}$. After the selection of $E$ and $q$ will be made, the value of the pre-Lagrangian for that specific selection will be identified with the actual Lagrangian density at $p$. To put (\ref{LJ3}) in a useful form, define two maps $f$, $g \colon S^3 \rightarrow S^2$, where here $S$ is any set, as follows:
\beq
f(a,b,c) = (a,b)\,, \qquad g(a,b,c) = (a,c)\,.
\eeq
With these definitions,
\bea
& &\int_{\alpha(p,q)} \dd^d r\, {\cal J}(F,f(p,q,r))
= \int_{\alpha(p,q)} \dd^d r\, {\cal J}(F,p,q)
= V(\alpha(p,q))\,{\cal J}(F,p,q) \kern20pt\\
\noalign{\noindent{and}}
& &\int_{\alpha(p,q)} \dd^d r\, {\cal J}(F,g(p,q,r))
= \int_{\alpha(p,q)} \dd^d r\, {\cal J}(F,p,r)\,,
\eea
and a pre-Lagrangian equivalent to (\ref{LJ3}) in the continuum is given by
\beq
{\cal L}_{\cal J}(F,E,p,q) = \frac{1}{V(\alpha(p,q))} \int_{\alpha(p,q)} \dd^d r\,
\big[{\cal J}\big(F,f(p,q,r)\big) + E\,{\cal J}\big(F,g(p,q,r)\big)\big]\,,
\eeq
because $V(\alpha(p,q))$ and $\tau^d$ are proportional, through a constant $k_d$ (see (\ref{kd}) below) that can be absorbed into $E$. In fact, since $\tau(p,q) = \tau_1$ will also be a constant, $V(\alpha(p,q))$ may be omitted, and in order to preserve time-reversal symmetry, we would like to replace ${\cal J}(F,f(p,q, r))$ and ${\cal J}(F,g(p,q,r))$ by ${\cal J}(F,f(p,q,r)) + {\cal J}(F,f(q,p,r))$ and ${\cal J}(F,g(p,q,r)) + {\cal J}(F,g(q,p,r))$, respectively, up to an unimportant factor of 2. Thus, the pre-Lagrangian can be written as
\bea
& &{\cal L}_{\cal J}(F,E,p,q) = \int_{\alpha(p,q)} \dd^d r\, \Big( {\cal J}\big(F,f(p,q,r)\big) + {\cal J}\big(F,f(q,p,r)\big) + \nonumber \\
& &\kern132pt+\ E\,\big[{\cal J}\big(F,g(p,q,r)\big) + {\cal J}\big(F,g(q,p,r)\big)
\big] \Big)\,. \kern20pt \label{Lsymm}
\eea
The Alexandrov-set function ${\cal J}(F,p,q)$ will be called a ``Lagrangian generator" for the pre-Lagrangian (\ref{Lsymm}). It is the basic object in the dynamics of the field $F$, and needs to be defined without reference to spacetime tensors or a differentiable structure on the underlying set.

More generally, in order to allow ${\cal L}_{\cal J}$ to include multiple integrals such as in (\ref{L1}) and (\ref{L2}), $f(p, q, r)$ and $g(p,q,r)$ should be generalized to $f(p,q, r_1, ..., r_n)$ and $g(p,q, r_1, ..., r_n)$. For the sake of completeness and convenience to the reader, our final definition of Lagrangian generator ${\cal J}$ will include both the prescription for going from the Lagrangian generator to the pre-Lagrangian ${\cal L}_{\cal J}(F,E,p,q)$ just discussed, and the transition from the pre-Lagrangian to the actual pointwise Lagrangian ${\cal L}_{\cal J}(F,x)$, discussed earlier in this section:
\medskip

\noindent {\em Definition:\/} Let $\cal F$ be the set of possible distributions of a set of field of interest. A Lagrangian generator is a triple $({\cal J},f,g)$ with ${\cal J} \colon {\cal F} \times S^n \rightarrow \mathbb{R}$, and $f$, $g \colon S^{2+m} \rightarrow S^n$, with $n \le 2+m$. The pre-Lagrangian corresponding to this Lagrangian generator is ${\cal L}_{{\cal J}} \colon {\cal F} \times \mathbb{R} \times S^2 \rightarrow \mathbb{R}$, given by 
\bea
& &{\cal L}_{\cal J}(F,E,p,q) = \sum_{p \prec r_i \prec q} \Big({\cal J}\big(F,f (p,q,r_1, ..., r_m)\big) + {\cal J}\big(F, f(q,p,r_1, ..., r_m)\big) + \nonumber \\
& &\kern112pt+\ E \big[ {\cal J}\big(F,g(p,q,r_1, ..., r_m)\big) + {\cal J}\big(F, g(q,p,r_1, ..., r_m)\big) \big] \Big)\,,\kern30pt
\eea
in the causal set case, with the sums replaced by integrals in the continuum case.
The ``variation" of this pre-Lagrangian in $\alpha(P,Q)$ is the real-valued function defined by
\beq
{\rm var}_{\tau_1}({\cal J},F,E,P,Q) = \max \{ {\cal L}_{\cal J}(F,E,p,q)
\mid P \prec p \prec q \prec Q\ \land\ \tau(p,q) = \tau_1 \}\,,
\eeq
and for any $x \in S$, the $(F, \tau_1, \tau_2)$-based fiducial neighborhood of $x$ is given by 
\bea
& &\alpha_{\tau_1,\tau_2}(F,x) = \{ (E,P,Q) \mid \tau(P,Q) = \tau_2\ \land\ \forall E',P',Q'\ {\rm with}\ \tau(P',Q') = \tau_2, \nonumber \\
& &\kern125pt {\rm var}_{\tau_1}({\cal L},F,E',P',Q')
\geq {\rm var}_{\tau_1}({\cal L},F,E,P,Q) \}\,.\kern40pt
\eea
Finally, the pointwise Lagrangian density corresponding to $\cal J$ is given by 
\beq
{\cal L}_{{\cal J},\tau_1,\tau_2}(F,x) = \frac{1}{\sharp\{\alpha_{\tau_1,\tau_2}(F,x)\}} \sum_{E,P,Q\in\alpha(x)} {\cal L}(F,E,P,Q)\,.
\eeq
Let us now see what this expression becomes for some important types of fields.

\section{Scalar Fields} 

The Lagrangian generator for a scalar field $\phi$ is given by $({\cal J},f,g)$, where 
\bea
& &{\cal J}(\phi,r,s) = \big(\phi(r) - \phi(s)\big)^2 - \half\,m^2 \phi^2(r) \\
& & f(r_1,r_2,r_3,r_4) = (r_1,r_2)\,, \qquad g(r_1,r_2,r_3,r_4) = (r_3,r_4)\,.
\eea
For reasons that will soon become apparent, $m$ is not the actual mass, although it is related to it. In fact, $m$ is assumed to be very small, of order $\tau$. 
The above expression implies that the pre-Lagrangian for a scalar field is given by 
\bea
& &{\cal L}(\phi,E,p,q) = \int \dd^d r\, \dd^d s\, \big({\cal J}(f(p,q,r,s))
+ E\, {\cal J}(g(p,q,r,s))\big) \\
& &=\ \int \dd^d r\, \dd^d s\, \big({\cal J}(\phi,p,q) + E\,{\cal J}(\phi,r,s)\big)
= {\cal J}(\phi,p,q)\, V^2(\alpha(p,q))
+ E \int\dd^d r\, \dd^d s\,{\cal J}(\phi,r,s)\,. \nonumber
\eea
By remembering that the volume of an $n$-dimensional ball is 
\beq
V({\rm ball}) = \frac{2 \pi^{n/2}}{n\,\Gamma(n/2)}\, r^n\,,
\eeq
we obtain, for the volume of an Alexandrov set in Minkowski space,
\beq
V(\alpha(p,q)) =  \frac{2 \pi^{(d-1)/2}}{(d-1)\, \Gamma((d-1)/2)} \int_{- \tau/2}^{\tau/2} \Big( \frac{\tau}{2} - |t| \Big)^{d-1} \dd t = k_d\, \tau^d\,,
\eeq
where
\beq
k_d = \frac{2\pi^{(d-1)/2}}{(d-1)\, \Gamma(2^{d-2}d\,(d-1))}\,. \label{kd}
\eeq
Substituting the above expression for volume, along with the Lagrangian generator $\cal J$, in the expression for $\cal L$, we obtain
\bea
& &{\cal L}(\phi,E,p,q) = k_d^2\, \tau^{2d}(p,q) \big[\big(\phi(q) - \phi(p)\big)^2
- \half\,m^2\, \big(\phi^2(p) + \phi^2(q)\big) \big] \nonumber\\
& &\kern70pt+\ E \int \dd^d r\, \dd^d s\, \big[\big(\phi(r) - \phi(s)\big)^2 - m^2 \phi^2(r)\big]\,.
\eea
We would now like to compute the Lagrangian density, if the spacetime is assumed to be flat Minkowskian and $\phi$ is assumed to be linear. 

Let's start from the mass term. It is given by
\beq
{\cal L}_m(\phi,E,p,q) = \frac{m^2\,k_d^2}{2}\, \tau^{2d}(p,q)\,
\big(\phi^2(p) + \phi^2(q)\big) + E\,m^2 \int \dd^d r\, \dd^d s\, \phi^2(r)\,.
\eeq
The above expression tells us that the leading order of the mass term is $m^2\,\tau^{2d}$. When we will get to the kinetic term, it will be shown that the order of magnitude that we are interested in is $\tau^{2d+2}$. Thus, if $m$ is assumed to be of the order of $\tau$, then the leading order is $\tau^{2d+2}$ which coincides with the leading order for the kinetic term. 

This means that, as far as the mass term is concerned, we can throw away all the higher-order terms. This can be done by using the approximation $\phi \approx \phi_0$, which tells us that
\bea
& &{\cal L}_m(\phi,E,p) \approx m^2\, k_d^2\, \tau^{2d}(p,q)\, \phi_0^2
+ E\,m^2\, \phi_0^2\, V^2(\alpha(p,q)) \nonumber \\ \noalign{\medskip}
& &\kern58pt =\ (1+E)\, m^2\, k_d^2\, \phi_0^2\, \tau^{2d}(p,q)\,.
\eea
Now let's look at the kinetic term. The linearity assumptions imply that in the integrand
\beq
\big(\phi(r) - \phi(s)\big)^2 = (r^\mu - s^\mu)\,(r^\nu - s^\nu)\,
\partial_\mu\phi\, \partial_\nu\phi\,.
\eeq
Consider a coordinate system in which the $t$ axis passes through $p$ and $q$, while the origin lies midway between these points. Denoting $\tau(p,q)$ by $\tau$, 
\beq
p = (-\tau/2, 0,0,0)\,, \qquad q = (\tau/2, 0, 0 ,0)\,.
\eeq
In this coordinate system, 
\beq
\int_{\alpha(p,q)} \dd^d r\, r^\mu = 0\,,
\eeq
since the above integrand is antisymmetric with respect to the center of the Alexandrov set. By slicing the Alexandrov set into $t=$ constant balls, we get 
\beq
\int_{\alpha(p,q)} (x^0)^2\, \dd^d r = \frac{2\pi^{(d-1)/2}}{(d-1)\,\Gamma((d-1)/2)}  \int_{-\tau/2}^{\tau/2}t^2 \Big( \frac{\tau}{2} - |t| \Big)^{d-1}\, \dd t
= I_{d0}\, \tau^{d+2}\,,
\eeq
where
\beq
I_{d0} = \frac{2\pi^{(d-1)/2}}{(d-1)\,\Gamma(2^{d-1}(d-1)\,d\,(d+1)\,(d+2))}\,.
\eeq
Furthermore, it can be shown that 
\beq
\int_{\alpha(p,q)} (x^k)^2\, \dd^d r = I_{d1}\, \tau^{d+2}\,,
\eeq
where, by cylindrical symmetry, the coefficient is the same for each $k$ and is given as 
\beq
I_{d1}=  \frac{2 \pi^{d/2-1}}{(d-2)\, \Gamma(d/2-1)}\,.
\eeq
Substituting these expressions into the integral in ${\cal L}_{\rm kin}$ we obtain
\bea
& &\int \dd^d r\, \dd^d s\, \big(\phi(r) - \phi(s)\big)^2
= \Big( \int \dd^d s \Big) \sum_{\mu=0}^{d-1} \dd^d r\, (r^\mu)^2
+ \Big( \int \dd^d r \Big) \sum_{\mu=0}^{d-1} \dd^d s\, (s^\mu)^2 \nonumber \\
& &\kern75pt=\ 2\,k_d\, \tau^{2d+2}\, \big[\big(I_{d0} + I_{d1} (d-1)\big)\,
(\partial_0 \phi)^2
- I_{d1}\,(d-1)\,\partial^\mu\phi\, \partial_\mu\phi \big]\,.\kern30pt
\eea
Thus, the kinetic term of the pre-Lagrangian is
\beq
{\cal L}_{\rm kin}(\phi,E,p,q) = k_d\, \tau^{2d+2}\, \big\{(\partial_0\phi)^2\, \big[k_d + 2\,E_d\,\big(I_{d0} + I_{d1}\,(d-1)\big)\big] - 2\, E_d\, I_{d1}\, (d-1)\, \partial^\mu\phi\, \partial_\mu\phi \big\}\,.
\eeq
Switching from the coordinate system in which the $t$ axis passes through $p$ and $q$ to an arbitrary one, the result becomes 
\bea
& &\kern-10pt{\cal L}_{\rm kin}(\phi,E,p,q)
= k_d\, \tau^{2d}(p,q)\, (q^\mu - p^\mu)\, (q^\nu - p^\nu)\,
\partial_\mu\phi\, \partial_\nu\phi\, \big[k_d + 2\,E_d\, \big(I_{d0} + I_{d1}\, (d-1)\big)\big] + \kern10pt \nonumber \\
\noalign{\smallskip}
& &\kern150pt -\ 2\, k_d\, E_d\, I_{d1}\, \tau^{2d+2}(d-1)\,
\partial^\mu\phi\, \partial_\mu\phi\,. \label{Lkin}
\eea
If the choice of points $p$ and $q$ varies with the constraints that both the midpoint $0$ between $p$ and $q$ as well as the Lorentzian distance between the two points are fixed, then ${\cal L}(\phi,E,p,q)$ undergoes a variation of order $\tau^{2d+2}$ due to the $(q^\mu-p^\mu)\, (q^\nu-p^\nu)\, \partial_\mu\phi\, \partial_\nu\phi$ term. The mass term, on the other hand, only gives variations to higher orders. Thus, if variations of orders higher than $\tau^{2d+2}$ are neglected, then the variation can be ``minimized", or in this case set to $0$, if the first term in (\ref{Lkin}) (depending on the ``tilting" of the axis of the Alexandrov set, while the second one is constant) vanishes, or
\beq
k_d + 2\,E_d\, \big(I_{d0} + I_{d1} (d-1)\big) = 0\,,
\eeq
which determines the value of $E_d$:
\beq
E_d = - \frac{k_d}{2\,\big(I_{d0} + I_{d1} (d-1)\big)}\,.
\eeq
Substituting this into the expression for the Lagrangian gives
\beq
{\cal L} = \frac{I_{d1}\,k_d^2\, (d-1)}{I_{d0} + I_{d1}\, (d-1)}\, \tau^{2d+2}\,
\partial^\mu\phi\, \partial_\mu\phi - m^2\, \phi^2\, k_d^2\, \tau^{2d}\,
\Big(1-\frac{k_d}{2\,\big(I_{d0} + I_{d1}\,(d-1)\big)} \Big)\,.
\eeq
This Lagrangian can be rewritten as 
\beq
{\cal L} = \hbar_d^2\, \frac{v_0}{2}\, \partial^\mu\phi_d\,
\partial_\mu\phi_d - \frac{v_0\, m_d^2}{2}\, \phi_d^2\,,
\eeq
where $v_0$ is the volume taken up by one point and 
\bea
& &\hbar_d \phi_d = \frac{k_d}{v_0}\, \tau^{d+1}
\sqrt{\frac{I_{d1}\,(d-1)}{I_{d0} + I_{d1}\,(d-1)}}\;\phi \\
& & m_d = \frac{k_d}{v_0}\,\tau^d
\sqrt{1-\frac{k_d}{2\,\big(I_{d0}+I_{d1}\,(d-1)\big)}}\;m\,.
\eea
In future sections other fields will be similarly scaled, but the coefficients will be different from field to field. At first this might seem wrong since the kinetic terms of all Lagrangians have the same coefficient $1$ in standard quantum field theory. But it can be easily shown that this difference does not amount to anything but a change of overall factor: 
\bea
& &\int [{\cal D} \phi_1 ...{\cal D} \phi_n]\,
\exp\big\{ \ii\, S(\phi_1)  + ... + \ii\, S(\phi_n) \big\} \\
& &= \rho_{1d}\, ...\, \rho_{nd} \int [{\cal D} \phi_{1d} ... {\cal D}\phi_{nd}]\,
\exp\big\{\ii\, S(\phi_{1d}/\rho) + ... + \ii\, S(\phi_{nd}/\rho)\big\}\,, \nonumber
\eea
where
\beq
\phi_{kd} = \rho_{kd}\, \phi_k\,.
\eeq

\section{Charged Scalar Field}

Consider a charged spin-0 particle, described by a set of complex scalar fields $\phi = (\phi_1, ..., \phi_n)$ coupled to a SU($n$) gauge field. (Notice that in this approach to the dynamics of matter fields in causal set theory, although it will be assumed that spacetime is discretized, the internal degrees of freedom will still have a continuous invariance group.) In the continuum, the dynamics of such a field can be described starting with the matter Lagrangian density
\beq
{\cal L}_{\rm m}(g_{\mu\nu},\phi,A_\mu;x) = \half\, |g|^{1/2}\,\big[g^{\mu\nu}\,
(D_\mu\phi)^{\dagger}\,(D_\nu\phi) - m^2\,\phi^{\dagger}\phi\big]\,, \label{Lm}
\eeq
where the gauge-covariant derivative is defined as usual by
$D_\mu\phi^a:= \partial_\mu\phi^a + \ii\,e\, A_\mu{}^a{}_b\,\phi^b$,
and $A_\mu = A_\mu{}^k\,T^k$ is the Lie-algebra-valued connection form representing the gauge field on a differentiable manifold. (Here, Latin indices $a$, $b$, ..., are Lie-algebra tensor indices, while $k$, $l$, ..., label elements of the basis $T^k$ of the Lie algebra.) In the causal set context, the scalar field will be simply replaced by a corresponding field defined at each causal set element, but to write down the action it is important to specify what variables will replace $A_\mu$.

A gauge field is defined in terms of holonomies, where the word ``holonomy" refers to the group transformation corresponding to the parallel transport of a Lie-algebra-valued field such as $\phi^a$ between two points $p$ and $q$. In a differentiable Lorentzian manifold $M$ (of dimension $d$), holonomy is defined as a function $a: M\times M \rightarrow {\rm TSU}(n)$, where TSU$(n)$ is the tangent bundle to $\SU(n)$, and consists therefore of all traceless $n \times n$ tensors. This map assigns to any two elements $p,\,q\in M$ the holonomy of $A_\mu$ along the geodesic segment $\gamma(p,q)$ connecting $p$ and $q$ in $M$, given by the path-ordered exponential
\beq
a(p,q) = \int_{\gamma(p,q)} A_\mu{}^k\,T^k\,\dd x^\mu\,, \label{holo}
\eeq
in terms of which the expression $D_\mu\phi(x)$ appearing in the scalar field Lagrangian arises from the leading-order term in the expansion of the expression $(1+a(x,y))\,(\phi(y) - \phi(x))$ .

This means that the causal set version of the charged scalar field Lagrangian can be obtained by making some simple substitutions in the one obtained in the previous section for the Klein-Gordon field. Thus, if the gauge field $a$ is assumed to be fixed and not subject to any Lagrangians, then the Lagrangian generator for a matter field $\phi \in \mathbb{R}^n$ interacting with $a$ is given by $({\cal J},f,g)$, where 
\beq
{\cal J}_{\rm sc}(\phi,a,r,s) = |\phi(r) - a(r,s)\, \phi(s)|^2
+ \textstyle{\frac18}\,m^2\,(\phi^*(r) + \phi^*(s))\, \big(\phi(r) + \phi(s)\big)
\eeq
and
\beq
f_{\rm sc}(r_1, r_2, r_3) = (r_1, r_3)\,,\qquad
g_{\rm sc}(r_1, r_2, r_3) = (r_1, r_2)\,.
\eeq
However, as discussed in the next section, $a$ itself is subject to a Lagrangian generator given by
\bea
& &{\cal J}_{\rm YM}(a,r_1,r_2,r_3)
= {\rm tr}[(a(r_1, r_2) + a(r_2,r_3) + a(r_3,r_1))^2]\,, \nonumber\\
\noalign{\smallskip}
& &f_{\rm YM}(r_1,r_2,r_3,r_4,r_5) = (r_3,r_4,r_5)\,, \nonumber\\
\noalign{\smallskip}
& &g_{\rm YM}(r_1,r_2,r_3,r_4,r_5) = (r_1,r_2,r_3)\,.
\eea
In order for the theory to possess an SU$(n)$ symmetry in the setup, these two Lagrangian generators are combined into one as $({\cal J}_{\rm tot},f,g)$, 
\bea
& &{\cal J}_{\rm tot}(a,r_1,r_2,r_3) = {\cal J}_{\rm sc}(\phi,a,r_1,r_2)
+ {\cal J}_{\rm YM}(a,r_1,r_2,r_3)\,, \nonumber \\
\noalign{\smallskip}
& &f_{\rm tot}(r_1,r_2,r_3,r_4,r_5) = f_{\rm YM}(r_1,r_2,r_3,r_4,r_5)\,, \nonumber \\
\noalign{\smallskip}
& &g_{\rm tot}(r_1,r_2,r_3,r_4,r_5) = g_{\rm YM}(r_1,r_2,r_3,r_4,r_5)\,.
\eea

\section{Yang-Mills field}

In this section, the main goal is to express the Yang-Mills Lagrangian density,
\beq
{\cal L}_{\rm YM}(g_{\mu\nu},A_\mu;x)
= \half\,|g|^{1/2}\,{\rm tr}(F_{\mu\nu}F^{\mu\nu})\,,
\eeq
in terms of the holonomies for the gauge field introduced in the previous section, as well as variables describing the geometry that are meaningful for causal set \cite {paper3}. Once this is done, the Lagrangian density can be easily rewritten in the discrete setting.

In the causal set context, a gauge field is defined as a map $a \colon S^2 \rightarrow {\rm TSU}(n)$. The Lagrangian generator for the gauge field is given by $({\cal J},f,g)$, where 
\beq
{\cal J}(a,r_1,r_2,r_3) = {\rm tr}[(a(r_1,r_2) + a(r_2,r_3) + a(r_3,r_1))^2]
\eeq
and 
\beq
f(r_1,r_2,r_3,r_4,r_5) = (r_1,r_2,r_3)\,,
\qquad g(r_1,r_2,r_3,r_4,r_5) = (r_3,r_4,r_5)\,.
\eeq
This means that in the case of Minkowski space the pre-Lagrangian is given by
\bea
& &\kern-25pt{\cal L}(a,E,p,q) = \int_{\tau(p,q)} \dd^d r\, \dd^d s\, \dd^d t\,
\big[{\cal J}(a,f(p,q,r,s,t)) + {\cal J}(a,f(q,p,r,s,t)) + \nonumber \\
& &\kern133pt +\ E\,\big({\cal J}(a,g(p,q,r,s,t)) + {\cal J}(a,g(q,p,r,s,t)\big)\big]
\nonumber \\
& &\kern-25pt = \int_{\tau(p,q)} \dd^d r\, \dd^d s\, \dd^d t\, \big[{\cal J}(a,p,q,r)
+ {\cal J}(a,q,p,r) + E\,({\cal J}(a,r,s,t) + {\cal J}(a,r,s,t)) \big]\;.
\eea
Integrating (trivially) over $s$ and $t$ in the first two terms, and using the fact that 
${\cal J}(a,p,q,r) = {\cal J}(a,q,p,r)$, we get that
\beq
{\cal L}(a,E,p,q) = 2\, \Big( V^2(\alpha(p,q)) \int_{\tau(p,q)} \dd^d r\,
{\cal J}(a,p,q,r) + E \int_{\tau(p,q)} \dd^d r\,\dd^d s\,\dd^d t\, {\cal J}(a,r,s,t)
\Big)\,.
\eeq
Assume that the gauge field is differentiable and reasonably well behaved. In particular, it is well-behaved-enough for $F_{\mu\nu}{}^k$ to be approximately constant in $\alpha(P,Q)$ whenever $(E,P,Q) \in \alpha_{\tau_1,\tau_2}(a,u)$ for some $u \in S$.  Assume for definiteness that the three points are spacelike related. Choose a coordinate system so that $r$ coincides with the origin, the $x$ axis points from $r$ to $s$, and the $y$ axis is perpendicular to the $x$ axis in the $r$-$s$-$t$ plane. Then in this coordinate system $a = (0,0,0,...)$, $b = (0,b^1,0,...)$, $c = (0,c^1,c^2,...)$. 

The flux of $F_{\mu\nu}{}^k$ through the interior of that triangle is expressed by the relationship
\beq
a(r,s)+ a(s,t)+ a(t,r) =  \half\, s^1\, t^2\, F_{12}{}^k\,T^k + ...
\eeq
This result generalizes to points at arbitrary locations, and can be written covariantly as
\beq
a(r,s)+a(s,t)+ a(t,r) =  \half\, F_{\mu\nu}{}^k\,T^k\,
(s^\mu - r^\mu)\, (t^\nu - r^\nu) + ...
\eeq
Recalling that, for SU($n$), tr$(T^kT^l) = C_2\, \delta_{kl}$, to leading order in the separation between points,
\beq
{\rm tr}[(a(r,s)+a(s,t)+a(t,r))^2] = \frac{C_2}{4}\, F_{\mu\nu}{}^k\, (s^\mu-r^\mu)\,
(t^\nu-r^\nu)\, F_{\rho\sigma}{}^k\, (b^\rho-a^\rho)\, (c^\sigma-a^\sigma)\,.
\label{trabc}
\eeq
Let's start from
\beq
\int_{\tau(p,q)} \dd^d r\, \dd^d s\, \dd^d t\, {\cal J}(a,r,s,t)\,.
\eeq
Expand the right-hand side of Eq (\ref{trabc}), and integrate term by term. Clearly any term with an odd number of powers of any variable will integrate to 0. Thus, the only terms that may potentially survive the integration are those of the form $r^\mu\, r^\nu\, r^\rho\, r^\sigma$ or quadratic terms in two of the three points. Simple counting of terms gives
\bea
& &\int_{p\prec r,s,t \prec q} \dd^dr\,\dd^ds\,\dd^dt\, {\rm tr}\big[(a(r,s)+a(s,t)+a(t,r))^2\big] \nonumber \\
& &= \frac{C_2}{4}\, F_{\mu\nu}{}^k\, F_{\rho\sigma}{}^k \int_{p\prec a,b,c\prec q} \dd^da\, \dd^db\,\dd^dc\, (s^\mu - r^\mu)\, (t^\nu - r^\nu)\, (s^\rho - r^\rho)\,
(t^\sigma - r^\sigma) \nonumber \\
& &= \frac{C_2}{4}\, \bigg[ 3\,V\sum_{k,\mu,\nu} (F_{\mu\nu}{}^k)^2
\bigg(\int_{\alpha(p,q)} \dd^da\, (r^\mu)^2 \bigg) \bigg(\int_{\alpha(p,q)} \dd^db\, (s^\nu)^2 \bigg) + \nonumber \\
& & \kern120pt -\ V^2 F_{\mu\nu}{}^k\, F_{\rho\sigma}{}^k\int_{\alpha(p,q)}
\dd^da\, r^\mu\, r^\nu\, r^\rho\, r^\sigma \bigg]\,, \label{intabc}
\eea
where $V$ is the volume of the Alexandrov set $\alpha(p,q)$.

The only terms of $F_{\mu\nu}{}^k\, F_{\rho\sigma}{}^k\, r^\mu\, r^\nu\, r^\rho\, r^\sigma$ that survive integration are the ones whose indices are pairwise equal. But if either $\mu = \nu$  or $\rho = \sigma$, then $F_{\mu\nu}{}^k = 0$ or $F_{\rho\sigma}{}^k = 0$, respectively, which would set the whole thing to 0. Thus, the only options are $\mu = \rho$, $\nu = \sigma$ and $\nu = \rho$, $\mu = \sigma$.  The antisymmetry of $F_{\mu\nu}{}^k$ then implies that these two cases are opposites of each other, which in turn implies that $F_{\mu\nu}{}^k\, F_{\rho\sigma}{}^k\,r^\mu\, r^\nu\, r^\rho\, r^\sigma = 0$. Thus, Eq (\ref{intabc}) becomes
\bea
& &\int_{p\prec r,s,t\prec q} \dd^dr\,\dd^ds\,\dd^dt\;
{\rm tr} \big[ (a(r,s)+a(s,t)+a(t,r))^2 \big]
\nonumber\\
& &= \frac{3\,VC_2}{4} \sum_{k,\mu,\nu} (F_{\mu\nu}{}^k)^2 \Big(\int_{\alpha(p,q)}
\dd^dr\, (r^\mu)^2\Big) \Big(\int_{\alpha(p,q)} \dd^ds\, (s^\nu)^2\Big) \nonumber\\
& &= \frac{3\, k_d\, C_2\, \tau^{3d+4}}{2}\, \sum_k \Big(J^0 J^1 \sum_{i=1}^{d-1}
(F_{i0}{}^k)^2 + (J^1)^2 \sum_{i<j} (F_{ij}{}^k)^2 \Big)\,, \label{intabc2}
\eea
where $J^\mu =\tau^{-d-2}\int_{\alpha(p,q)} \dd^dx\, (x^\mu)^2$, or in other words
\bea
& &J^0
= \frac{2\pi^{(d-1)/2}}{2^d\, (d-1) \, d\,(d+1)\,(d+2)}\, \Gamma((d-1)/2) \\
& & J^1 = ... = J^{d-1} 
= \frac{2\pi^{d/2-1}}{2^{d+1}\, (d-2)\, d\,(d+2)}\, \Gamma((d-2)/2)
\int_{-\pi/2}^{\pi/2}(\cos\theta)^d\,\dd\theta\,.
\eea
Now let's move to the second integral, $\int_{\alpha(p,q)} \dd^dx\,{\rm tr}[(a(p,x)+a(x,q)+a(q,p))^2]$, where $p \prec q$ are the endpoints of the Alexandrov set. Rewriting Eq (\ref{trabc}) in terms of the points $p$, $x$, and $q$ gives
\bea
& &{\rm tr}\big[ (a(p,x)+a(x,q)+f(q,p))^2 \big] \\ \noalign{\medskip}
& &=\ \fourth\,C_2\,F_{\mu\nu}{}^k\,(p^\mu-x^\mu)\,(q^\nu-x^\nu)\,F_{\rho\sigma}{}^k\, (p^\rho-x^\rho)\,(q^\sigma-x^\sigma)\,. \nonumber
\label{trApB}
\eea
Again this can be expanded and integrated term by term. There are several conditions each term has to meet, in order for its integral not to vanish. First of all, it needs to contain an even number of factors of $x$. Secondly, as was shown before, for symmetry reasons
\beq
F_{\mu\nu}{}^k\, F_{\rho\sigma}{}^k \int_{\alpha(p,q)}\, \dd^dx\,
\, x^\mu\, x^\nu\, x^\rho\, x^\sigma = 0\,.
\eeq
Finally, $F_{\mu\nu}\, p^\mu p^\nu = F_{\mu\nu}\, q^\mu q^\nu = 0$,
and the identities $p = (-\frac{\tau}{2},0,0,0)$ and $q = (\frac{\tau}{2},0,0,0)$ imply $F_{\mu\nu}\, p^\mu q^\nu = -F_{\mu\nu}\, p^\mu p^\nu = 0$. The only terms in Eq (\ref{trApB}) that do {\em not\/} vanish for any of the above reasons are 
\bea
& &F_{\mu\nu}{}^k\, F_{\rho\sigma}{}^k\, p^\mu\, x^\nu\, p^\rho\, x^\sigma\,,\qquad
F_{\mu\nu}{}^k\, F_{\rho\sigma}{}^k\, p^\mu\, x^\nu\, x^\rho\, q^\sigma\,,
\nonumber\\ \noalign{\medskip}
& &F_{\mu\nu}{}^k\, F_{\rho\sigma}{}^k\, x^\mu\, q^\nu\, p^\rho\, x^\sigma\,,\qquad
F_{\mu\nu}{}^k\, F_{\rho\sigma}{}^k\, x^\mu\, q^\nu\, x^\rho\, q^\sigma\,.
\nonumber
\eea
Plugging in the coordinate values of $p$ and $q$,
each of the above four expressions evaluates to 
$\fourth\,\tau^2\, F_{\mu0}{}^k\, F_{\rho0}{}^k\, x^\mu\, x^\rho$.
In order for this not to be an odd function, $\mu = \rho$ has to hold, and
in order for $F_{\mu 0}$ to be non-zero $\mu \not= 0$ has to hold. Thus,
this becomes $\fourth\,\tau^2\, (F_{i0}{}^k)^2\, (x^i)^2$ and, since
there are four such terms, the integral becomes
\bea
& & \int_{\alpha(p,q)} \dd^dx\;{\rm tr}\big[ (a(p,x)+a(x,q)+a(q,p))^2 \big]
\nonumber \\
& &= \fourth\,C_2\, \tau^2 \sum_{i=1}^{d-1} \int_{\alpha(p,q)} \dd^dx\,(F_{i0}{}^k)^2\, (x^i)^2 = \fourth\,C_2\, \tau^{d+4} J^1 \sum_{i=1}^{d-1} (F_{i0}{}^k)^2\,,
\eea
where, based on rotational symmetry, $J^1 = ... = J^{d-1}$ was used. 

Substituting this into the original expression for the pre-Lagrangian, and doing some basic algebra, the latter becomes
\beq
{\cal L}(a,E,p,q) = \tau^{3d+4}\, C_2\, k_d\, J^1\, \Big( (3\,E J^0 + \fourth\,k_d) \sum(F^k_{i0})^2 + 3\,E J^1 \sum_{i<j} (F^k_{ij})^2 \Big)\,.
\eeq
In order to get rid of the variation that results from different choices of Alexandrov sets, we would like the above expression to be relativistically invariant; in other words, we would like it to be proportional to $F^{\mu \nu} F_{\mu \nu}$. This means
\beq
3\,E_d\, J^0 + \half\,k_d = -3\,E_d J^1\,,
\eeq
which implies
\beq
E_d = - \frac{k_d}{6\,(J^0 +J^1)}\,.
\eeq
Substituting this expression for $E_d$ gives the Lagrangian
\beq
{\cal L}(a,p,q) = -\frac{k_d^2\,(J^1)^2\,C_2}{4\,(J^0+J^1)}\,
\tau_2^{3d+4}\, F^{\mu\nu} F_{\mu\nu}\,,
\eeq
which can be expressed as
\beq
{\cal L}(a,p,q) = - \frac{v_0}{4}\, F_d{}^{\mu\nu}\, F_{d\mu\nu}\,,
\eeq
where $v_0$ is the volume taken up by a single point if $F_d{}^{\mu\nu}$ is defined as
\beq
F_d{}^{\mu\nu} = k_d\, J^1\, F^{\mu\nu}
\sqrt{\frac{C_2\,\tau^{3d+4}}{v_0\,(J^0 + J^1)}}\,.
\eeq
This means that
\beq
A_d{}^\mu = k_d\, J^1\, A^\mu \sqrt{ \frac{C_2\, \tau^{3d+4}}{v_0\, (J^0 + J^1)}}\,.
\eeq
From this we can determine how the charge changes from dimension to dimension. At a first glance, since we haven't yet defined a Lagrangian for fermions, we are only ready to talk about the charge of bosonic fields. However, a simple symmetry consideration allows us to overcome this barrier and include fermionic charges in the discussion.  Whether a field is bosonic or fermionic, we would like to be able to say 
\beq
\partial_\mu \rightarrow \partial_\mu + e\,A_\mu\,.
\eeq
We would also like to be able to say 
\beq
\partial_\mu \rightarrow \partial_\mu + e_d\, A_{d\mu}\,.
\eeq
This means that, regardless whether the field in question is bosonic or fermionic, and regardless of any other properties of the field (such as mass) it should satisfy
\beq
e\,A^\mu = e_d\, A^\mu_d\,.
\eeq
This immediately implies that, both for bosons and fermions,
\beq
e_d = \frac{e}{k_d J^1} \sqrt{\frac{v_0\, (J^0 + J^1)}{C_2\, \tau^{3d+4}}}\,.
\eeq

\section{Gravitational field}

The Lagrangian generator for gravity is the triple $({\cal J},f,g)$, defined simply by
\beq
{\cal J}(r,s,t) = \begin{cases}
(8\pi G)^{-1} & \hbox{if}\ r \prec t \prec s\\
0& \hbox{otherwise}
\end{cases}\;,
\qquad f(r,s,t,u) = (r,s,t)\,, \qquad g(r,s,t,u) = (r,t,u)\,.
\eeq
This means that the pre-Lagrangian is given by
\beq
{\cal L} = \int \dd^d r\, \dd^d s\, \sqrt{|\det g(r)|}\,\sqrt{|\det g(s)|}\,
\big[{\cal J}\big(f(p,q,r,s)\big) + E\,{\cal J}\big(g(p,q,r,s)\big)\big]\,.
\eeq
Since ${\cal J}$ is a constant inside a certain domain and is $0$ outside, the above integrals amount to restricting $r$ and $s$ to certain domains. The first term is ${\cal J}(p,q,r)$, thus its restriction is $p \prec r \prec q$. That statement, of course, is trivial, which means that the first term has no restriction at all. On the other hand, the second term is ${\cal J}(p,r,s)$, thus it has a restriction $p \prec r \prec s \prec q$. Taking this into account, the pre-Lagrangian becomes
\bea
& &\kern-20pt{\cal L} = \frac{1}{8\pi G} \Big(\int \dd^d r\, \dd^d s\,
\sqrt{|\det g(r)|}\, \sqrt{|\det g(s)|}
+ E \int_{p \prec r \prec s \prec q} \dd^d r\, \dd^d s\,
\sqrt{|\det g(r)|}\,\sqrt{|\det g(s)|} \Big) \nonumber\\
& &\kern-9pt=\ \big(V(\alpha(p,q))\big)^2
+ E \int_{\alpha(p,q)} \dd^d s\, \sqrt{|\det g(s)|}\, V(\alpha(p,s))\,.
\eea 
One can parametrize the interior of the Alexandrov set $\alpha(p,q)$ with normal geodesic coordinates around $p$. Suppose that Greek indices $\mu$, $\nu$ are used for coordinates satisfying
\beq
g_{\mu\nu}(p) = \eta_{\mu\nu}\,,
\eeq
but which are otherwise arbitrary, not necessarily geodesic. Then one can define the normal geodesic coordinates derived from the above in the following way: 
\beq
r^{\bar\mu} = \eta^{\mu\nu}\, \partial_\nu\gamma_{pr}\big\vert_p\,
\Big(\int_{\gamma_{pr}} g_{\mu\nu} \dd x^\mu\, \dd x^\nu\Big)^{1/2}\,,
\eeq
where $\gamma_{ab}$ denotes the geodesic segment connecting $a$ and $b$. 
From this, it is straightforward to see that the following equation is satisfied exactly:
\beq
\tau(p,r) = \eta_{\bar\mu\bar\nu}\, r^{\bar\mu}\, r^{\bar\nu}\,.
\eeq
It should be noticed, however, that the above is true only if one of the two points is $p$, while
\beq
\tau(r,s) \neq \eta_{\bar\mu\bar\nu}\, (s^{\bar\mu}
-r^{\bar\mu})\, (s^{\bar\nu} - r^{\bar\nu})\,.
\eeq
However, a different causal relation, $\prec_p$, will be introduced in addition to the already existing one, $\prec$. While $\prec$ matches flat spacetime expectations only if the pair of points of interest includes $p$, $\prec_p$ does so for arbitrary pairs of points:
\beq
a \prec_p b\quad \Leftrightarrow\quad \eta_{\bar\mu\bar\nu}\, (b^{\bar\mu}
-a^{\bar\mu})\, (b^{\bar\nu} - a^{\bar\nu}) \geq 0\,.
\eeq
At the same time, we will retain the original causal relation, $\prec$, for which the above is not true:
\beq
{}\prec{} \neq {}\prec_p{}\,.
\eeq
Let's introduce the following notation:
\bea
& &J^+(a) = \{r \succ a \}\,, \qquad J_p^+(a) = \{r \succ_p a \}\,, \nonumber \\
& & J^-(a) = \{r \prec a \}\,, \qquad J_p^-(a) = \{r \prec_p a \} \\
& &\kern-34pt \alpha(a,b) = J^+(a) \cap J^-(b)\,, \qquad \alpha_p(a,b) = J_p^+(a) \cap J_p^-(b)\,,
\eea
and define $V_p(a,b)$, not to be confused with $V(\alpha_p(a,b))$, as follows
\beq
V_p(a,b) = k_d\, \big(\eta_{\mu\nu}\, (b^\mu-a^\mu)\, (b^\nu-a^\nu)\big)^{d/2}\,.
\eeq
It is important to notice that 
\beq
{\rm vol}(\alpha_p(a,b)) \neq V_p(a,b)\,,
\eeq
because the definition of $V_p$ neglects the $|g|^{1/2}$ factor in the volume element. But, for our purposes, $V_p$ as defined above is the simplest one to use. On the other hand, $V(a,b)$ is defined as the actual volume of the Alexandrov set,
\beq
V(a,b) = {\rm vol}(\alpha(a,b))\,.
\eeq
Then, the integral of the volume can be expanded as follows
\bea
& &\int_{\alpha(p,q)} \dd^d r \,\sqrt{|\det g(r)|}\, V(\alpha(p,r)) \nonumber\\
& &= \int_{\alpha_p(p,q)} \dd^d r\, V_p(p,q)
+ \Delta_{1d}(p,q) + \Delta_{2d}(p,q) + \Delta_{3d}(p,q) + O(\tau^{2d+4})\,,
\eea
where $\Delta_{1d}(p,q)$ is the correction coming from the mismatch between $\alpha(p,q)$ and $\alpha_p(p,q)$, $\Delta_{2d}(p,q)$ is the correction due to the error in $V(\alpha(p,r))$, and $\Delta_3$ is the correction due to the use of $\dd^d r$ instead of $\dd^d r \sqrt{(-1)^{d-1} \det g}$ for the volume element. These corrections are formally defined as follows: 
\bea
& &\Delta_{1d}(p,q) = \int_{(\alpha(p,q) \setminus \alpha_p(p,q))} \dd^d r\, V_p(p,r)
- \int_{\alpha_p(p,q) \setminus \alpha(p,q))} \dd^d r\, V_p(p,r) \\
& &\Delta_{2d}(p,q)= \int_{\alpha_p(p,q)} \dd^d r\, (V(p,r) - V_p(p,r)) \\
& &\Delta_{3d}(p,q) = \int_{\alpha_p(p,q)} \dd^d r\,
\big({\textstyle\sqrt{(-1)^{d-1} \det g}-1}\big)\,.
\eea
Whenever any of these three correction terms is computed the other two are neglected, since the ``correction of the correction" is of order $\tau^{2d+4}$, while the calculation is performed to order $\tau^{2d+2}$. Thus, the shape of the Alexandrov set is assumed to be unchanged in the calculation of $\Delta_{2d}$ and $\Delta_{3d}$, $\sqrt{(-1)^{d-1} \det g}$ is dropped in the calculation of $\Delta_{1d}$ and $\Delta_{2d}$, and the correction to $V(\alpha(p,x))$ is neglected in the calculation of $\Delta_{1d}$ and $\Delta_{3d}$.

Let's start by computing $\Delta_{1d}$. In normal coordinates, the light cone of $p$ is not deformed by curvature, while the light cone of $q$ still is,
\beq
J^+(p) = J_p^+(p)\,, \qquad J^-(q) \neq J_p^-(q)\,.
\eeq
Substitution of above into simple set theory algebra gives
\bea
& &\alpha_p(p,q) \setminus \alpha(p,q) = J_p^+(p) \cap (J_p^-(q) \setminus J^-(q)) \subset J_p^-(q) \setminus J^-(q) \\
\noalign{\smallskip}
& &\alpha(p,q) \setminus \alpha_p(p,q) = J_p^+(p) \cap (J^-(q) \setminus J_p^-(q)) \subset J^-(q) \setminus J_p^-(q)\,.
\eea
Assuming that $J_p^-(q)$ and $J^-(q)$ are very close to each other, the above implies that most of the contribution to $\Delta_{1d}$ comes from the vicinity of $J_p^-(q)$. So, let's evaluate $V_p(p,r)$ for $r \in J_p^-(q)$. In other words, assume that 
\beq
\sum(r^{\bar{k}})^2 = (\tau - r^{\bar0})^2\,.
\eeq
This implies that 
\beq
\eta_{\bar\mu\bar\nu}\, r^{\bar\mu}\, r^{\bar\nu}
= (r^{\bar0})^2 - (\tau - r^{\bar0})^2 = \tau\, (2r^{\bar0}
- \tau)\,,
\eeq
which, using the fact that
\beq
2\,r^{\bar0} - \tau = r^{\bar0} - (\tau - r^{\bar0})
= r^{\bar0}- {\textstyle\sqrt{\sum(r^{\bar{k}})^2}}\;,
\eeq
becomes
\beq
\eta_{\bar\mu\bar\nu}\, r^{\bar\mu}\, r^{\bar\nu}
= \tau_0\,\Big(r^{\bar0}-{\textstyle\sqrt{\sum(r^{\bar{k}})^2}}\Big)\,.
\eeq
Now define two functions $\chi_1$ and $\chi_2$ as follows:
\bea
& &\kern-20pt\chi_1(r) = \int_{\alpha(r,q)} \dd^ds\,k_d\, (\eta_{\bar\mu\bar\nu}\, s^{\bar\mu}\, s^{\bar\nu})^{d/2} - \int_{\alpha_p(r,q)} \dd^ds\,k_d\, (\eta_{\bar\mu\bar\nu}\, s^{\bar\mu}\, s^{\bar\nu})^{d/2}\\
& &\kern-20pt\chi_2(r) = \int_{\alpha(r,q)} \dd^ds\, k_d\, \Big[\tau_0\, \Big(s^{\bar0} - {\textstyle\sqrt{\sum(s^{\bar{k}})^2}}\Big) \Big]^{d/2} - \int_{\alpha_p(r,q)} \dd^ds\,k_d\, \Big[\tau_0\, \Big(s^{\bar0} - {\textstyle\sqrt{\sum(s^{\bar k})^2}}\Big) \Big]^{d/2}.\kern30pt
\eea
As a consequence of the fact that $J^+(p) = J_p^+(p)$, we get
\beq
\chi_1(p) = \chi_2(p)\,,
\eeq
and it is easy to see that 
\beq
\chi_1(q) = \chi_2(q) = 0\,.
\eeq
Therefore 
\beq
\Delta_{1d} = \chi_1(p) = - (\chi_2(q) - \chi_2(p))
= -\int_0^{\tau(p,q)} \dd\tau\,\frac{\dd\chi_2(r(\tau))}{\dd\tau}\,,
\eeq
where $r(\tau)$ is defined as a point on $\gamma_{pq}$ whose distance to $r$ is $\tau$:
\beq
\gamma_{pr(\tau)} \subset \gamma_{pq}\,, \qquad l(\gamma_{pr(\tau)}) = \tau\,.
\eeq
This immediately implies that 
\beq
\delta\tau > 0 \Rightarrow r(\tau) \prec r(\tau + \delta\tau)\,.
\eeq
Furthermore, in normal coordinates
\beq
r^{\bar\mu}(\tau) = \tau\, V^{\bar\mu}\,,
\eeq
where $V^{\bar\mu}$ is a tangent vector to $\gamma_{pq}$ at $p$. The fact that the latter is timelike implies that
\beq
\delta\tau > 0\quad \Rightarrow\quad r(\tau) \prec_p r(\tau + \delta\tau)\,.
\eeq
In order to compute $\dd\chi_2(r(\tau))/\dd\tau$ notice that, schematically, 
\bea
& &\delta\chi_2(r) = \int_{S_1} f - \int_{S_2} f - \int_{S_3} f + \int_{S_4} f = \int_{S_1 \setminus S_2} f - \int_{S_2 \setminus S_1} f - \int_{S_3 \setminus S_4} f + \int_{S_4 \setminus S_3}f \nonumber \\
& &=\ \int_{(S_1 \setminus S_2) \setminus (S_3 \setminus S_4)} f - \int_{(S_3 \setminus S_4) \setminus (S_1 \setminus S_2)} f - \int_{(S_3 \setminus S_4) \setminus (S_4 \setminus S_3)}f + \int_{(S_4 \setminus S_3) \setminus (S_3 \setminus S_4)}f\,,\kern10pt
\eea
where
\beq
S_1 = \alpha(r(\tau + \delta\tau),q)\,, \quad
S_2 = \alpha_p(r(\tau + \delta\tau),q)\,, \quad
S_3 = \alpha(r(\tau),q)\,, \quad
S_4 = \alpha_p(r(\tau),q)\,.\quad
\eeq
Substituting both $r(\tau) \prec r(\tau + \delta\tau)$ and $r(\tau) \prec_p r(\tau + \delta\tau)$ and using simple set-theory algebra, one obtains
\bea
& &\alpha(r(\tau),q) \setminus \alpha_p(r(\tau),q) \setminus (\alpha(r(\tau
+ \delta\tau),q) \setminus \alpha_p(r(\tau + \delta\tau),q)) \nonumber \\
& &= (J^+(r(\tau)) \setminus J^+(r(\tau + \delta\tau))) \cap (J^-(q) \setminus \alpha_p(r(\tau,q))) \subset J^+(r(\tau)) \setminus J^+(r(\tau + \delta\tau))
\kern40pt\\
\noalign{\medskip}
& &\alpha_p(r(\tau),q) \setminus \alpha(r(\tau),q) \setminus (\alpha_p(r(\tau + \delta\tau),q) \setminus \alpha(r(\tau + \delta\tau),q)) \nonumber \\
& &= (J_p^+(r(\tau)) \setminus J_p^+(r(\tau + \delta\tau))) \cap (J_p^-(q) \setminus \alpha(r(\tau,q))) \subset J_p^+(r(\tau)) \setminus J_p^+(r(\tau + \delta\tau))\\
\noalign{\medskip}
& &(\alpha(r(\tau + \delta\tau),q) \setminus \alpha_p(r(\tau + \delta\tau), q)) \setminus (\alpha(r(\tau), q) \setminus \alpha_p(r(\tau),q)) \nonumber \\
& &= \alpha(r(\tau + \delta\tau),q) \cap (J_p^+(r(\tau)) \setminus J_p^+(r(\tau + \delta\tau))) \subset J_p^+(r(\tau)) \setminus J_p^+(r(\tau + \delta\tau)) \\
\noalign{\medskip}
& &(\alpha_p(r(\tau + \delta\tau),q) \setminus \alpha(r(\tau + \delta\tau), q)) \setminus (\alpha_p(r(\tau),q) \setminus \alpha(r(\tau), q)) \nonumber \\
& &= \alpha_p(r (\tau + \delta\tau),q) \cap (J^+(r(\tau)) \setminus J^+(r(\tau
+ \delta\tau))) \subset J^+(r(\tau)) \setminus J^+(r(\tau + \delta\tau))\,.
\eea
Thus, all four integrals are performed either over a subset of $J^+(r(\tau)) \setminus J^+(r(\tau + \delta\tau))$ or over a subset of $J_p^+(r(\tau)) \setminus J_p^+(r(\tau + \delta\tau))$. In either case, the range of integration is in the vicinity of the lightcone of $r(\tau)$.

Now let $s$ be an arbitrary point in that region. Remembering that $\tau$ denotes the distance from $p$ to the point of intersection of that region with the geodesic $\gamma_{pq}$, and is not to be confused with the distance from $p$ to a floating point, we have  
\bea
& &\kern-20pt s \in \big[J^+(r(\tau)) \setminus J^+(r(\tau + \delta\tau))\big]
\cup \big[J_p^+(r(\tau)) \setminus J_p^+(r(\tau + \delta\tau))\big] \nonumber \\
& &\kern-20pt \Rightarrow {\textstyle\sqrt{\sum(s^{\bar{k}})^2}} = s^{\bar0} - \tau + O(\tau^2) \Rightarrow k_d\, \Big[\tau_0 \Big(s^{\bar0} - {\textstyle\sqrt{\sum(s^{\bar{k}})^2}} \Big)\Big]^{d/2} = k_d\, (\tau\tau_0)^{d/2} + O(\tau^{d+2})\,.\kern30pt
\eea
Thus, all integrals reduce to integrations over a constant:
\bea
& &\kern-12pt\int_{S_1} f\, \dd^d x - \int_{S_2} f\, \dd^d x
- \int_{S_3} f\, \dd^d x + \int_{S_4} f\, \dd^d x \nonumber \\
& &\kern-12pt=\  \Big(\int_{(S_1 \setminus S_2) \setminus (S_3 \setminus S_4)} - \int_{(S_3 \setminus S_4) \setminus (S_1 \setminus S_2)} - \int_{(S_3 \setminus S_4) \setminus (S_4 \setminus S_3)} + \int_{(S_4 \setminus S_3) \setminus (S_3 \setminus S_4)}\Big)\, f\, \dd^d x \nonumber \\
& &\kern-12pt=\ (\tau \tau_0)^{d/2}\, \Big(\int_{(S_1 \setminus S_2) \setminus (S_3 \setminus S_4)} - \int_{(S_3 \setminus S_4) \setminus (S_1 \setminus S_2)}
- \int_{(S_3 \setminus S_4) \setminus (S_4 \setminus S_3)} + \int_{(S_4 \setminus S_3 \setminus (S_3 \setminus S_4)}\Big)\, \dd^d x \nonumber \\
& &\kern-12pt=\ (\tau \tau_0)^{d/2}\, \Big( \int_{S_1} \dd^d x - \int_{S_2}\dd^d x - \int_{S_3} \dd^d x + \int_{S_4} \dd^d x \Big) \nonumber \\
& &\kern-12pt=\ (\tau \tau_0)^{d/2}\, \big(V(S_1)- V(S_2) - V(S_3) + V(S_4)\big)\,.\vphantom{\int}
\eea
Now, $V(S_1)$, $V(S_2)$, $V(S_3)$ and $V(S_4)$ can be read off from reference \cite{GibSol} as follows:
\bea
& &\kern-20pt V(S_1) = \big[k_d + (A_d\,Rg_{\mu\nu} + B_d\,R_{\mu\nu})\,
\big(q^\mu - r^\mu\, (\tau + \delta\tau)\big)\, \big(q^\nu - r^\nu\,
(\tau + \delta\tau)\big) \big]\, \tau^d(r(\tau + \delta\tau),q) \nonumber\\
\noalign{\smallskip}
& &\kern-20pt V(S_2) = k_d\, \tau^d\big(r(\tau + \delta\tau), q\big) \nonumber\\
\noalign{\smallskip}
& &\kern-20pt V(S_3) = (k_d + (A\,Rg_{\mu\nu} + B\,R_{\mu\nu})\,(q^\mu - r^\mu(\tau))\,
(q^\nu - r^\nu(\tau)))\, \tau^d(r(\tau),q) \nonumber\\
\noalign{\smallskip}
& &\kern-20pt V(S_4) = k_d\, \tau^d(r(\tau), q)\,.
\eea
Since the coordinate system is defined in terms of geodesics coming out of $p$, all of these geodesics, including $\gamma_{pq}$ are, by definition, straight lines in the chosen coordinate system. Therefore, it can be assumed that $\gamma_{pq}$ coincides with the $t$ axis, which simplifies the above equations:
\bea
& & V(S_1) = k_d\, (\tau_1 - \tau - \delta\tau)^d
+ (A_d\,R + B_d\,R_{00})\, (\tau_1 - \tau - \delta\tau)^{d+2}\\
\noalign{\smallskip}
& & V(S_2) = k_d\, (\tau_1 - \tau - \delta\tau)^d \\
\noalign{\smallskip}
& & V(S_3) = k_d\, (\tau_1 - \tau)^d + (A_d\,R+B_d\,R_{00})\, (\tau_1 - \tau)^{d+2}\\
\noalign{\smallskip}
& & V(S_4) = k_d\, (\tau_1 - \tau)^d\,.
\eea
This implies that 
\bea
& &\kern-15pt\chi_2 (r(\tau + \delta\tau)) - \chi_2 (r(\tau))
= (\tau_1 \tau)^{d/2}\big(V(S_1) - V(S_2) - V(S_3) + V(S_4)\big) \\
& &\kern107pt=\ (A_d\,R + B_d\,R_{00})\, (\tau_1 \tau)^{d/2}\, \frac{\dd}{\dd\tau}\,
(\tau_1 - \tau)^{d+2}\,\delta\tau + O(\tau^{d+2}\,(\delta\tau)^2)\,,\kern30pt\nonumber
\eea
which in turns implies that
\beq
\frac{\dd\chi_2(r(\tau))}{\dd\tau} = (A_d\,R + B_d\,R_{00})\, (\tau_1\tau)^{d/2}
\,\frac{\dd}{\dd\tau}\, (\tau_1 - \tau)^{d+2}\,.
\eeq
Thus, 
\beq
\Delta_{1d} = \chi_1(p) - \chi_1(q) = \chi_2(p) - \chi_2(q)
= -(A_d\,R+B_d\,R_{00}) \int_0^{\tau_1} \dd\tau\, (\tau_1\tau)^{d/2}\,
\frac{\dd}{\dd\tau}(\tau_1-\tau)^{d+2}\,.
\eeq
The binomial expansion of $(\tau_1 - \tau)^{d+2}$ gives
\beq
\Delta_{1d} = -(A_d\,R+B_d\,R_{00}) \int_0^{\tau_1} \dd\tau\, (\tau_1\tau)^{d/2}\, \frac{\dd}{\dd\tau} \bigg(\sum_{k=0}^{d+2} (-1)^k {{d+2}\choose k}\, \tau_1^{d+2-k}
\,\tau^k \bigg)\,.
\eeq
Evaluating the derivative and combining it with the $\tau^{d/2}$ factor we find that
\beq
\Delta_{1d} = -(A_d\,R+B_d\,R_{00}) \int_0^{\tau_1} \dd\tau\, \tau_1^{d/2}  \bigg( \sum_{k=0}^{d+2} (-1)^k {{d+2} \choose k}\, \tau_1^{d+2-k}\, k\,
\tau^{d/2+k-1} \bigg)\,.
\eeq
Finally, integration of this last expression gives
\beq
\Delta_{1d} = (A_d\,R+B_d\,R_{00})\, \tau_1^{2d+2}
\sum_{k=0}^{d+2} (-1)^k {{d+2} \choose k} \frac{k}{d/2+k}\,.
\eeq
Now let's compute $\Delta_{2d}$. We will go back to our usual notation, where $r$ (which will be denoted by $x$ in order to avoid conflict with $r$ being a radius of a ball) is an arbitrary element of $\alpha(p,q)$, and it is no longer assumed to lie on $\gamma_{pq}$. As was stated earlier, the $\Delta_{2d}$ correction is due to the error of computing the volume of $\alpha(p,x)$, where $x$ is far away from the boundary of $\alpha(p,q)$. Since that correction is already of order $\tau^{d+2}$, its integral over $\alpha(p,q)$ is of order $\tau^{2d+2}$, while its integral over the ``corrections" to the shape of $\alpha(p,q)$ is of order $\tau^{2d+4}$. For this reason, the latter term will be neglected, and it will be assumed that the shape of $\alpha(p,q)$ has not been affected by curvature, while the shapes of $\alpha(p,x)$ has. Furthermore, in light of the geodesic coordinates, while the corrections to the volume of $\alpha(p,x)$ are not neglected, the flat space equation for the distance will be used in computing them. Thus, the equation for $\Delta_{2d}$ becomes
\beq
\Delta_{2d} = A_d\, R \int_{\alpha_p(p,q)} \dd^d x\,
(x^{\bar\mu}\, x_{\bar\mu})^{1+d/2} + B_d\, R_{\bar\mu\bar\nu}
\int_{\alpha_p(p,q)} \dd^d x\, x^{\bar\mu}\, x^{\bar\nu}\,
(x^{\bar\rho}\, x_{\bar\rho})^{d/2}\,.
\eeq
By using the cylindrical symmetry, this becomes
\bea
& &\Delta_{2d} = A_d\, R \int_{\alpha_p(p,q)} \dd^d\overline{x}\,
(x^{\bar\mu} x_{\bar\mu})^{1+d/2} 
+ B_d\, R_{\bar0\bar0} \int_{\alpha_p(p,q)} \dd^d\overline{x}\,
(x^{\bar0})^2\, (x^{\bar\rho}\, x_{\bar\rho})^{d/2}\;+ \nonumber \\
& &\kern33pt +\ B_d\,\big(\sum R_{kk}\big) \int_{\alpha_p(p,q)} \dd^d\overline{x}\, (x^{\bar1})^2\, (x^{\bar\rho}\, x_{\bar\rho})^{d/2}\,,
\eea
where, due to cylindrical symmetry, $(x^{\bar{k}})^2$ was replaced with
$(x^{\bar1})^2$. 

From cylindrical symmetry, 
\beq
\int_{\alpha_p(p,q)} \dd^d\overline{x}\, (x^{\bar\rho}\, x_{\bar\rho})^{1 + d/2}
= \int_{\alpha_p(p,q)} \dd^d\overline{x}\, (x^{\bar0})^2\,
(x^{\bar\rho}\, x_{\bar\rho})^{d/2} - (d-1) \int \dd^d\overline{x}\,
(x^{\bar1})^2\, (x^{\bar\rho}\, x_{\bar\rho})^{d/2}\,.
\eeq
This can be used to get rid of the integral involving $(x^{\bar1})^2$, and get
\beq
\Delta_2 = \tau^{2d+2} \Big[R \Big( \Big(A_d + \frac{B_d}{d-1} \Big) H_{d,2,d/2} \Big) + \frac{R_{00}}{d-1} \big(d\,B_d\,H_{d,2,d/2} - B_d\,H_{d,0,1+d/2}\big)\Big]\,,
\eeq
where
\beq
H_{d,i,j} = \frac{1}{\tau^d} \int_{\alpha_p(p,q)} \dd^d\overline{x}\,
(x^{\bar0})^i\, (x^{\bar\mu}\, x_{\bar\mu})^j
\eeq
and
\beq
H_{d,i,j+1/2} = \frac{1}{\tau^d} \int_{\alpha_p(p,q)} \dd^d\overline{x}\,
(x^{\bar0})^i\, (x^{\bar\mu}\, x_{\bar\mu})^{j+d/2}\,;
\eeq
here $j$ can be either an integer or a half-integer. 

Let us now compute these coefficients. In the calculations that follows we will treat the above integrands simply as functions, and ``forget" that their source is a curvature. If we imagine slicing $\alpha_p(p,q)$ into $\overline{t} =$ constant balls and then slicing each ball into $\overline{r} =$ constant spheres, it is easy to see that 
\bea
& &\int_{\alpha_p(p,q)} \dd^d\overline{x}\, (x^{\bar0}
- p^{\bar0})^i\, \big((x^{\bar\mu} - p^{\bar\mu})\,
(x_{\bar\mu} - p_{\bar\mu})\big)^{j/2} \dd^d x \\
& &=\ \frac{2 \pi^{(d-1)/2}}{ \Gamma ((d-1)/2)} \Big( \int_0^{\tau/2} \dd\overline{t}\;\bar{t}^i \int_0^t \dd\overline{r}\; \overline{r}^{d-2}\, (\overline{t}^2 - \overline{r}^2)^{j/2} + \int_{\tau/2}^{\tau} \dd\overline{t}\; \overline{t}^i \int_0^{\tau - \overline{t}} \dd\overline{r}\; \overline{r}^{d-2} (\overline{t}^2 - \overline{r}^2)^{j/2} \Big)\,.\kern8pt \nonumber
\eea
By changing variables to
\beq
\overline{u} = \frac{\overline{t}}{\tau}\,, \qquad
\overline{s} = \frac{\overline{r}}{\overline{t}}\;,
\eeq
the above expression becomes
\bea
& &\kern-30pt\int_{\alpha_p(p,q)} \dd^d\overline{x}\, (x^{\bar0} - p^{\bar0})^i \big((x^{\bar\mu} - p^{\bar\mu})\, (x_{\bar\mu} - p_{\bar\mu})\big)^{j/2} \nonumber \\
& &\kern-30pt=\ \frac{2 \pi^{(d-1)/2}}{\Gamma((d-1)/2)}\, \tau^{i+j+d}\, \bigg[ \int_0^{1/2}
\dd\overline{u}\, \Big( \overline{u}^{i+j+d-1} \int_0^1 \dd\overline{s}\; \overline{s}^{d-2} (1-\overline{s}^2)^{j/2} \Big) +\nonumber \\
& &\kern85pt+\ \int_{1/2}^1 \dd\overline{u}\, \Big( \overline{u}^{i+j+d-1} \int_0^{1/u-1} \dd\overline{s}\; \overline{s}^{d-2}\, (1-\overline{s}^2)^{j/2} \Big) \bigg]\,.
\eea
Since the limits of integration in the first term are constants, that  term can be represented as a product of two separate integrals. After evaluating the $u$-integral, the expression becomes 
\bea
& &\int_{\alpha_p(p,q)} \dd^d\overline{x}\, (x^{\bar0}
- p^{\bar0})^i\, \big((x^{\bar\mu} - p^{\bar\mu})\,
(x_{\bar\mu} - p_{\bar\mu})\big)^{j/2} \nonumber \\
& &= \frac{2\pi^{(d-1)/2}}{\Gamma((d-1)/2)}\,\tau^{i+j+d}\, \bigg[\frac{1}{(i+j+d)\,2^{i+j+d}} \int_0^1 \dd\overline{s}\; \overline{s}^{d-2}\, (1-\overline{s}^2)^{j/2} + \\
& &\kern110pt+\ \int_{1/2}^1 \dd\overline{u}\, \Big(\overline{u}^{i+j+d-1} \int_0^{1/u-1} \dd\overline{s}\; \overline{s}^{d-2}\, (1-\overline{s}^2)^{j/2}\Big) \bigg]\,. \nonumber
\eea
From now on the calculation splits into four cases: even and odd $d$ and even and odd $j$. From the original intentions of the calculation it is clear that whenever $d$ is odd $j$ is also add, and viceversa. So only these two cases need be considered.
\medskip

\noindent {\em Case 1:\/} Both $d$ and $j$ are even. Since $j$ is even, denote it as
\beq
j = 2\,h\,.
\eeq
Expanding $(1-\overline{s}^2)^h$ binomially, the integral becomes
\bea
& &\int_{\alpha_p(p,q)} \dd^d\overline{x}\, (x^{\bar0} - p^{\bar0})^i\,
\big((x^{\bar\mu} - p^{\bar\mu})\, (x_{\bar\mu} - p_{\bar\mu})\big)^{j/2} \nonumber \\
& &=\ \frac{2\pi^{(d-1)/2}}{\Gamma((d-1)/2)}\, \tau^{i+j+d} \sum_{k=0}^h (-1)^k {h \choose k} \bigg[ \frac{1}{(i+2h+d)\,2^{i+2h+d}} \int_0^1 \dd\overline{s}\;\overline{s}^{d-2+2k} + \\
& &\kern182pt+\ \int_{1/2}^1 \dd\overline{u}\, \Big( \overline{u}^{i+2h+d-1} \int_0^{\frac{1}{\overline{u}}-1}\, \dd\overline{s}\; \overline{s}^{d-2+2k} \Big) \bigg]\,. \nonumber
\eea
After evaluating the $\overline{s}$ integrals this becomes
\bea
& &\kern-30pt\int_{\alpha_p(p,q)} \dd^d\overline{x}\, (x^{\bar0}
- p^{\bar0})^i \big((x^{\bar\mu} - p^{\bar\mu})\,
(x_{\bar\mu} - p_{\bar\mu})\big)^{j/2} \nonumber \\
& &\kern-30pt=\ \frac{2 \pi^{(d-1)/2}}{ \Gamma ((d-1)/2)}\, \tau^{i+j+d} \sum_{k=0}^h (-1)^k {h \choose k} \bigg[ \frac{1}{2^{i+2h+d}\, (i+2h+d)\, (d-1+2k)} + \nonumber \\
& &\kern150pt+\ \int_{1/2}^1 \dd\overline{u}\, \Big( \overline{u}^{i+2h+d-1}\,
\frac{(1/\overline{u}-1)^{d-1+2k}}{d-1+2k} \Big) \bigg]\,.
\eea
After pulling out $d-1+2k$ and expanding out $(1/\overline{u}-1)^{d-1+2k}$, that becomes
\bea
& &\kern-20pt\int_{\alpha_p(p,q)} \dd^d\overline{x}\, (x^{\bar0}
- p^{\bar0})^i\, \big((x^{\bar\mu} - p^{\bar\mu})\, (x_{\bar\mu} - p_{\bar\mu})\big)^{j/2} \nonumber \\
& &\kern-20pt=\ \frac{2\pi^{(d-1)/2}}{\Gamma((d-1)/2)}\, \tau^{i+j+d} \sum_{k=0}^h \frac{(-1)^k}{d-1+2k} {h \choose k} \bigg( \frac{1}{2^{i+2h+d}\,(i+2h+d)} + \\
& &\kern185pt+\ \sum_{l=0}^{d-1+2k} (-1)^l {d-1+2k \choose l} \int_{1/2}^1
\dd\overline{u}\; \overline{u}^{i+2h-2k+l} \bigg)\,. \kern10pt\nonumber
\eea
Evaluating the integral that is left this becomes
\beq
\int \dd^d\overline{x}\, (x^{\bar0} - p^{\bar0})^i\, \big((x^{\bar\mu} - p^{\bar\mu})\, (x_{\bar\mu} - p_{\bar\mu})\big)^{j/2}  = H_{dih}\, \tau^{i+j+d}\,,
\eeq
where
\bea
& &H_{dih}= 2\pi^{(d-1)/2}\, \sum_{k=0}^h \frac{(-1)^k}{d-1+2k} {h \choose k} \bigg( \frac{1}{2^{i+2h+d}\,(i+2h+d)}\;+ \\
& &\kern178pt+\ \sum_{l=0}^{d-1+2k} (-1)^l\, {d-1+2k \choose l}
\,\frac{1-(\frac12)^{1+2h-2k+l+1}}{i+2h-2k+l+1} \bigg)\,. \nonumber
\eea

\noindent {\em Case 2:\/} Both $d$ and $j$ are odd. Since $j$ is odd, it will be replaced with
\beq
j = 2\,h+1\,.
\eeq
Then the binomial expansion tells us that
\beq
(1-\overline{s}^2)^j = \sqrt{1-\overline{s}^2}\, \sum_{k=0}^j {h \choose k}
(-1)^k\,\overline{s}^{2k}\,,
\eeq
which means that the original integral can be rewritten as
\bea
& &\kern-30pt\int_{\alpha_p(p,q)} \dd^d\overline{x}\, (x^{\bar0} - p^{\bar0})^i \big((x^{\bar\mu} - p^{\bar\mu})\, (x_{\bar\mu} - p_{\bar\mu})\big)^{j/2}
= \frac{2 \pi^{(d-1)/2}}{\Gamma((d-1)/2)}\,\tau^{2h+1+i+d}\;\times\nonumber\\
& &\kern30pt\times\ \sum_{k=0}^h {h \choose k} (-1)^k
\bigg[ \frac{1}{(i+2h+1+d)\,2^{i+2h+1+d}}
\int_0^1\dd\overline{s}\; \overline{s}^{d-2+2k} \sqrt{1-\overline{s}^2}\ + \nonumber\\
& &\kern115pt+\ \int_{1/2}^1 \dd\overline{u} \Big( \overline{u}^{i+2h+d} \int_0^{\frac{1}{\overline{u}}-1} \dd\overline{s}\; \overline{s}^{d-2+2k} \sqrt{1-\overline{s}^2}  \Big) \bigg]\,.
\eea
Since $d$ is odd, so is $d-2+2k$. Thus,
\beq
d-2+2k = 2a+1\,,
\eeq
where
\beq
a = \frac{d-3+2k}{2}\,.
\eeq
Thus, the integral of interest is $\int \overline{s}^{2a+1}\,
\sqrt{1-\overline{s}^2}\, \dd\overline{s}$.

By using
\beq
\overline{s} = \sin \overline{\theta}
\eeq
the integral becomes 
\beq
\int \overline{s}^{2a+1}\, \sqrt{1-\overline{s}^2}\, \dd\overline{s}
= \int \sin^{2a+1}\overline{\theta}\,\cos\overline{\theta}\,\dd(\sin\overline{\theta})
= \int \sin^{2a+1}\overline{\theta}\,\cos^2\overline{\theta}\,\dd\overline{\theta}\,.
\eeq 
Combining one of the $\sin \overline{\theta}$ factors with $\dd\overline{\theta}$ gives 
\beq
\int \overline{s}^{2a+1}\,\sqrt{1-\overline{s}^2}\,\dd\overline{s}
= - \int \sin^{2a}\overline{\theta}\,\cos^2\overline{\theta}\,
\dd(\cos\overline{\theta})\,.
\eeq
Expanding $\sin^{2a} \overline{\theta}$ as
\beq
\sin^{2a} \overline{\theta} = (1 - \cos^2 \overline{\theta})^a
= \sum_{b=0}^a (-1)^b {a \choose b} \cos^{2b} \overline{\theta}
\eeq
the above integral becomes 
\bea
& &\int \dd\overline{s}\;\overline{s}^{2a+1}\,\sqrt{1-\overline{s}^2}
= - \sum_{b=0}^a\, (-1)^b\, {a \choose b} \int \dd(\cos\overline{\theta}) \cos^{2b+2}\overline{\theta} \\
& &\kern97pt= - \sum_{b=0}^a (-1)^b {a \choose b}
\frac{\cos^{2b+3}\overline{\theta}}{2b+3}
\eea
By substituting $s= \sin \theta$ this becomes 
\beq
\int \dd\overline{s}\;\overline{s}^{2a+1}\,\sqrt{1-\overline{s}^2} = - \sum_{b=0}^a \frac{(-1)^b}{2b+3} {a \choose b} (1-\overline{s}^2)^{b+3/2}\,.
\eeq
Substituting $2a+1=d-2+2k$ we obtain
\beq
\int \dd\overline{s}\;\overline{s}^{d-2+2k}\,\sqrt{1-\overline{s}^2}
= - \sum_{b=0}^{k+\frac{d-3}{2}} \frac{(-1)^b}{2b+3} {k+\frac{d-3}{2} \choose b}
(1-\overline{s}^2)^{b+ 3/2}\,.
\eeq
Thus, the original integral becomes
\bea
& &\int_{\alpha_p(p,q)} \dd^d\overline{x}\, (x^{\bar0} - p^{\bar0})^i\, \big((x^{\bar\mu} - p^{\bar\mu})\, (x_{\bar\mu} - p_{\bar\mu})\big)^{h+1/2} \nonumber \\
& &=\ \frac{2\pi^{(d-1)/2}}{\Gamma((d-1)/2)} \sum_{k=0}^h \bigg\{ (-1)^k {h \choose k} \sum_{b=0}^{k+\frac{d-3}{2}} \frac{(-1)^b}{2b+3}{k+\frac{d-3}{2} \choose b} \times \nonumber \\
& &\kern15pt\times\ \bigg[ \frac{1}{(i+2h+1+d)\, 2^{i+2h+1+d}} + \int_{1/2}^1 \dd\overline{u}\, \overline{u}^{i+2h+d} \Big( \Big(\frac{2}{\overline{u}} - \frac{1}{\overline{u}^2} \Big)^{b+ 3/2}-1 \Big) \bigg] \bigg\}\,.
\eea
By evaluating the $(-1)$ term of the integral over $\overline{u}$, and also pulling $\overline{u}$ out of the denominators, the expression becomes
\bea
& &\int_{\alpha_p(p,q)} \dd^d\overline{x}\, (x^{\bar0} - p^{\bar0})^i\,
\big((x^{\bar\mu} - p^{\bar\mu})\, (x_{\bar\mu} - p_{\bar\mu})\big)^{h+1/2}
\nonumber \\
& &=\ \frac{2 \pi^{(d-1)/2}}{\Gamma((d-1)/2)} \sum_{k=0}^h \Big[ (-1)^k {h\choose k} \sum_{b=0}^{k+\frac{d-3}{2}} \frac{(-1)^b}{2b+3} {k+\frac{d-3}{2} \choose b}
\Big( \frac{1}{(i+2h+1+d)\, 2^{i+2h+1+d}} + \nonumber \\
& &\kern100pt-\ \frac{1-(\frac{1}{2})^{i+2h+d+1}}{i+2h+d+1}  + \int_{1/2}^1 \dd\overline{u}\, \overline{u}^{i+2h+d-2b-3} (2\,\overline{u}-1)^{b+3/2} \Big) \Big]\,.
\eea
By expanding $(2\,\overline{u}-1)^{b+1}$ we get
\bea
& &\int_{1/2}^1\dd\overline{u}\;\overline{u}^{i+2h+d-2b-3}\,
(2\,\overline{u}-1)^{b+3/2} \\
& &= \sum_{c=0}^b (-1)^c {b+1 \choose c} 2^{b-c+1} \int_{1/2}^1 \dd\overline{u}\; \overline{u}^{i+2h+d-b-c-2}\, \sqrt{2\,\overline{u}-1}\,. \nonumber
\eea
By setting
\beq
\overline{v}= \sqrt{2\,\overline{u}-1}
\eeq
this becomes 
\bea
& &\int_{1/2}^1\dd\overline{u}\;\overline{u}^{i+2h+d-2b-3}
\,(2\,\overline{u}-1)^{b+3/2} \\
& &= \sum_{c=0}^b (-1)^c {b+1 \choose c}\, 2^{b-c+1}
\int_0^1 \dd\overline{v}\; \overline{v}^2
\Big( \frac{\overline{v}^2 +1}{2} \Big)^{i+2h+d-b-2-c}\,.
\eea
We can carry out the $\overline v$ integration if we expand the power of the binomial $(\overline{v}^2+1)$, and we get
\bea
& &\int_{1/2}^1 \dd\overline{u}\, \overline{u}^{i+2h+d-2b-3}\,
(2\,\overline{u}-1)^{b + 3/2} \\
& & = \sum_{c=0}^b \bigg( (-1)^c {b+1 \choose c}\, 2^{2b-i-2h-d+3}
\sum_{e=0}^{i+2h+d-b-2-c} {1+2h+d-b-2-c \choose e}\, \frac{1}{2e+3}\bigg)\,. \nonumber
\eea
Substituting these into the original integral we obtain
\beq
\int_{\alpha_p(p,q)} \dd^d\overline{x}\, (x^{\bar0} - p^{\bar0})^i\, \big((x^{\bar\mu} - p^{\bar\mu})\, (x_{\bar\mu} - p_{\bar\mu})\big)^{h+1/2} = H_{d,i,h+1/2}\, \tau^{2h+i+d+1}\,,
\eeq
where
\bea
& &\kern-25pt H_{d,i,h+1/2} = \frac{2 \pi^{(d-1)/2}}{\Gamma((d-1)/2)}
\;\sum_{k=0}^h \bigg\{ (-1)^k  {h \choose k} \sum_{b=0}^{k+(d-3)/2}
\frac{(-1)^b}{2b+3} {k + \frac{d-3}{2} \choose b} \times \nonumber \\
& &\times\ \bigg[ \frac{1}{(i+2h+d+1)\,2^{i+2h+d+1}}
- \frac{1-(\frac12)^{i+2h+d+1}}{i+2h+d+1}\ + \\
& &+\ \sum_{c=0}^b \bigg( (-1)^c {b+1 \choose c} 2^{2b-i-2h-d+3}
\sum_{e=0}^{i+2h+d-b-2-c} {1+2h+d-b-2-c \choose e} \frac{1}{2e+3}
\nonumber \bigg) \bigg] \bigg\}\,.
\eea
Finally, let's compute $\Delta_3$. Writing the metric in normal coordinates, 
\beq
g_{\bar\mu\bar\nu} = \eta_{\bar\mu\bar\nu} - \third\, R_{\bar\mu\bar\rho\bar\nu\bar\sigma}\, x^{\bar\rho} \,x^{\bar\sigma}\,,
\eeq
we find that
\beq
\sqrt{|\det g|} = 1 - \sixth\, R_{\bar\rho\bar\sigma}\, x^{\bar\rho}\, x^{\bar\sigma}\,.
\eeq
Thus,
\beq
\Delta_{3d} = - \frac{k_d}{6}\, R_{\bar\rho\bar\sigma}
\int\dd^d x\,x^{\bar\rho}\, x^{\bar\sigma}\, \big((x^{\bar\alpha}
- p^{\bar\alpha})\, (x_{\bar\alpha}-p_{\bar\alpha})\big)^{d/2}\,.
\eeq
As was done with $\Delta_2$, the correction to the correction term will be neglected, which means that integration is performed over $\alpha_p(p,q)$ instead of $\alpha(p,q)$ and no correction term is introduced to the $V(\alpha(p,x))$ when the expression $((x^{\bar\alpha} - p^{\bar\alpha})\, (x_{\bar\alpha} - p_{\bar\alpha}))^d$ was used. 

Substituting the above into the expression for $\Delta_{3d}$ gives
\beq
\Delta_{3d} = \frac{k_d}{6}\, R_{\bar\rho\bar\sigma} \int_{\alpha_p(p,q)}
\dd^d\overline{x}\, x^{\bar\rho}\, x^{\bar\sigma}\, \big((x^{\bar\alpha}
- p^{\bar\alpha})\, (x_{\bar\alpha} - p_{\bar\alpha})\big)^{d/2}\,.
\eeq
By cylindrical symmetry, $(x^k)^2$ in the above integral can be replaced with $(x^1)^2$ which means
\bea
& &\Delta_{3d} = \frac{k_d}{6} \bigg[ R_{\bar0\bar0} \int_{\alpha_p(p,q)}
\dd^d\overline{x}\, (x^{\bar0})^2\, \big((x^{\bar\alpha} - p^{\bar\alpha}\big)\,
(x_{\bar\alpha} - p_{\bar\alpha}))^{d/2} + \nonumber \\
& &\kern50pt-\ \Big(\sum R_{\bar{k}\bar{k}} \Big) \int_{\alpha_p(p,q)} \dd^d \overline{x}\, (x^{\bar1})^2\, \big((x^{\bar\alpha} - p^{\bar\alpha})\,
(x_{\bar\alpha} - p_{\bar\alpha})\big)^{d/2} \bigg]\,.
\eea
Again, by cylindrical symmetry,
\bea
& &\int_{\alpha_p(p,q)} \dd^d\overline{x}\, \big((x^{\bar\alpha} - p^{\bar\alpha})\, (x_{\bar\alpha} - p_{\bar\alpha}) \big)^{1+ d/2}
= \int_{\alpha_p(p,q)} \dd^d\overline{x}\, (x^{\bar0})^2\, \big((x^{\bar\alpha} - p^{\bar\alpha})\, (x_{\bar\alpha} - p_{\bar\alpha})\big)^{d/2} + \nonumber\\
& &\kern120pt-\ (d-1)\int_{\alpha_p(p,q)} \dd^d\overline{x}\, (x^{\bar1})^2\, \big((x^{\bar\alpha} - p^{\bar\alpha})\, (x_{\bar\alpha} - p_{\bar\alpha}) \big)^{d/2}\,,
\eea
which allows us to express the integral involving $(x^{\bar1})^2$ in terms of integrals involving $(x^{\bar0})^2$. Substituting this expression into the expression for $\Delta_{3d}$ and doing some simple algebra gives
\beq
\Delta_{3d} = - \frac{k_d}{6\,(d-1)}\,\tau^{2d+2}\, \big[R\, (H_{d,0,1 + d/2}
- H_{d, 2, d/2}) + R_{00}\, (d\,H_{d,2,d/2} - H_{d,0,1 + d/2})\big]\,.
\eeq
Adding $\Delta_{1d}$, $\Delta_{2d}$ and $\Delta_{3d}$ together, the total correction becomes
\beq
\Delta_d = \Delta_{1d} + \Delta_{2d} + \Delta_{3d}
= \tau^{2d+2}\, (C_d\,R+D_d\,R_{00})\,,
\eeq
where
\beq
C_d = A_d \sum_{k=0}^{d+2} (-1)^k {d+2 \choose k} \frac{k}{d/2+k}
+ H_{d,0,d/2+1} \Big( A_d + \frac{B_d}{d-1} - \frac{k_d}{6\,(d-1)} \Big)
+ \frac{H_{d,2,d/2}}{d-1} \Big( \frac{k_d}{6} -B_d \Big)
\eeq
and
\beq
D_d = B_d \sum_{k=0}^{d+2} (-1)^k {d+2 \choose k} \frac{k}{d/2+k}
+ \frac{1}{d-1} \Big[ \Big( B_d - \frac{k_d}{6} \Big) d\,H_{d,2,d/2}
+ \Big( \frac{d\,k_d}{6} - B_d \Big) H_{d,0,1+d/2} \Big]\,.
\eeq
Now, as we were computing the corrections to the integral, we almost forgot the main term! Here it is:
\beq
\int_{\alpha_p(p,q)} \dd^d x\, k_d\, \big((x^{\bar\mu} - p^{\bar\mu})\,(x_{\bar\mu} - p_{\bar\mu})\big)^{d/2} = k_d\, \tau^{2d}\,H_{d,0,d/2}\,.
\eeq
This means that the total integral is 
\beq
\int_{\alpha(p,q)} \dd^d x\, V(\alpha(p,x))
= k_d\, H_{d,0,d/2}\, \tau^{2d} + (C_d\, R + D_d\, R_{00})\, \tau^{2d+2}\,.
\eeq
At the beginning of this section it was shown that the gravitational pre-Lagrangian is given by 
\beq
{\cal L}(\prec,E,p,q) = \frac{1}{4\pi G} \Big( V^2(\alpha(p,q))
+ E \int_{\alpha(p,q)} \dd^d r\, {\textstyle\sqrt{(-1)^{d-1}\det g}}\,V(p,r)\Big)\,.
\eeq
By substituting the above expression for the integral, as well as 
\beq
V(\alpha(p,q)) = k_d\, \tau^d + \tau^{d+2}\, (A_d\, R + B_d\, R_{00})\,,
\eeq
we obtain 
\bea
& &{\cal L}(\prec,E,p,q) = \tau^{2d}\,k_d\, (k_d + E\, H_{d,0,d/2}) \\
\noalign{\smallskip}
& &\kern75pt+\ \tau^{2d+2} \big(R\,(2\,k_d\,A_d + E\,C_d) + R_{00}\, (2\,k_d\,B_d + E\,D_d)\big)\,. \nonumber
\eea
Thus, in order to minimize variations, we have to get rid of the $O(\tau^{2d})$ contribution, which comes form the $R_{00}$ term. Thus, $E_d$ is selected in such a way that the coefficient of $R_{00}$ vanishes:
\beq
2\,k_d\,B_d + E_d\,D_d = 0 \Rightarrow E_d = -2\,k_d\, B_d/D_d\,.
\eeq

Finally, substituting this into the expression for the gravitational Lagrangian gives
\beq
{\cal L}(\prec,p,q) = \frac{1}{4\pi G} \bigg[\tau^{2d}\,k_d^2 \Big(1- \frac{2B_d}{D_d}\, H_{d,0,d/2} \Big) + 2\,k_d\, R\, \tau^{2d+2} \Big(A_d - \frac{B_d\, C_d}{D_d} \Big) \bigg]\,. \label{grav}
\eeq
As long as $d$ is fixed, the first term in this Lagrangian is constant, and in fact has the form of a cosmological constant; its role, however, needs to be studied in more detail, since it appears to have a very large value, while a different type of argument based on causal sets by R Sorkin \cite{cc} leads to a value of the right order of magnitude for the cosmological constant. The second term in (\ref{grav}) is proportional to the scalar curvature $R$, as expected, and can be rewritten in the more familiar form 
\beq
{\cal L}_{\rm EH} = \frac{v_0}{8\pi\, G_d}\, R\,,
\eeq
where $v_0$ is the volume taken up by a single point of a causal set and $G_d$ is given by 
\beq
G_d = G\, \frac{v D_d}{2\,k_d\,\tau_2^{2d+2}\, (A_d\,D_d - B_d\,C_d)}\,.
\eeq 

Regarding the use of the above gravitational Lagrangian, it should be noted that variations in the partial order $\prec$ are much more general than the variations in the metric that lead to the Einstein equation in the continuum version of the theory. One can speculate that, while causal-set variations that stay within the class of manifoldlike ones will still lead to an effective continuum gravitational field equation, based on the second term of (\ref{grav}), the first term in that expression suggests that other types of variations may lead to dynamical restrictions on how manifoldlike ``classical causal sets" are allowed to be, and particularly to conditions on their dimensionality. Such topics are beyond the scope of this contribution.

\section{Conclusion}

In this work, Lagrangians for the main types of bosonic fields \cite{paper5} have been successfully defined for arbitrary causal sets. The machinery we introduced allowed us to derive otherwise complicated-looking causal-set Lagrangians from simple Alexandrov-set based Lagrangian generators, and we showed that, if the relevant fields are defined on a Lorentzian manifold and the Alexandrov sets are small enough that we can use the appropriate slow-variation approximation for the fields inside them, the Lagrangian densities coincide with the ones used in standard field theory. For similar work on fermionic fields, see reference \cite{paper6}. We should point out here that, in light of the non-linearity of the procedure for selecting an Alexandrov set, while Lagrangians for different fields can be added, the same is not true for Lagrangian generators. Thus, for a set of fields with individual Lagrangian generators ${\cal J}_1$, ..., ${\cal J}_n$, we can formally write
\beq
{\cal J} = \{Ê{\cal J}_1 , ..., {\cal J}_n \}\;,\qquad
{\cal L}_{\cal J} = \sum\nolimits_i {\cal L}_{{\cal J}_i}\;.
\eeq

This line of research inherits the same problem that other approaches to quantum field theory have to face: if we treat gravity like all other fields, then a path integral over it implies an integration over all possible topologies and geometries, thus leaving us with none of the background information that is needed to define propagators or solve any other problems in ordinary quantum field theory. While this issue has been addressed in other approaches \cite{Rovelli}, our belief is that in our context these should be addressed by an appropriate theory of quantum measurement that accommodates general relativistic covariance and includes gravity as one of the ``measured" fields. A formulation of that theory is beyond the scope of this work.

Another difficulty is that even if the topology was not an issue (say, we are dealing with a toy model of propagation of non-gravitational fields in fixed gravitational background), there are simply too many degrees of freedom to integrate over. In the case of regular quantum field theory this issue is addressed by visualizing spacetime as a regular cubic lattice, which allows one to perform the integrals over ``all lattice points at once". In the case of a general causal set, however, such structure is absent, which means that one has to separately integrate over the fields at every single point in spacetime, one by one. While numerical simulations on the computer might make it possible to do so in toy models of causal sets consisting of only a few points, this becomes a problem for more realistic models. There is a way to make oneself feel better about it: The Einstein equation does not have many exact solutions either. Nevertheless, in order to bring causal set theory to the level of the Einstein equation, there have to be some simple cases for which there is a solution, such as for example a field with a very large gradient which assures a specific selection of Alexandrov sets. This, however, is still work for a future. 

The above problems are made much worse by the fact that causal set theory claims to be a candidate for a quantum theory of gravity, which implies that it should be precise enough to talk about the gravitational field of elementary particles: a precision much higher than the one of any other existing theory. Thus, reducing it to crude approximations might well defeat the very purpose causal set theory is there in the first place.  

Finally, the fact that the selection of Alexandrov sets is non-linear implies that the superposition principle does not work on microscopic scale. This might also put under question other principles that are based on linearity, such as the law of conservation of energy. In the case of smooth differentiable behavior, since these Lagrangians do reduce to the ones observed in regular quantum field theory, all the known laws, including the law of conservation of energy, should hold. They can, however, be violated once the behavior of fields is no longer smooth (such as, for instance, near the big bang). On the positive side, the existence of a source of non-conservation might have verifiable cosmological consequences. 


\bigskip

\section*{References}
\begin{thebibliography}{77}

\bibitem{CS} Bombelli L, Lee J, Meyer D and Sorkin R D 1987 Space-time as a causal set {\em Phys. Rev. Lett.} {\bf59} 521.

\bibitem{SorkinCausal} Sorkin R D 2003 Causal sets: discrete gravity
{\em Lectures on Quantum Gravity\/} eds A Gomberoff and D Marolf
(New York: Springer) ({\em Preprint\/} gr-qc/0309009)

\bibitem{Dowker} Dowker F 2005 Causal sets and the deep structure of spacetime {\em 100 Years of Relativity -- Space-time Structure: Einstein and Beyond\/} ed A Ashtekar (World Scientific) ({\em Preprint\/} arXiv:gr-qc/0508109)

\bibitem{Henson} Henson J 2009 The causal set approach to quantum gravity {\em Approaches to Quantum Gravity: Toward a New Understanding of Space, Time and Matter\/}
ed D Oriti (Cambridge University Press) ({\em Preprint\/} gr-qc/0601121)

\bibitem{SGD} Rideout D and Sorkin R 2000 Classical sequential growth
dynamics for causal sets {\em Phys. Rev.} D {\bf61} 024002 ({\em Preprint\/} gr-qc/9904062)

\bibitem{paper1} Sverdlov R and Bombelli L 2009 Gravity and matter in causal set theory {\em Class. Quantum Grav.} {\bf26} 075011 ({\em Preprint\/} 0801.0240)

\bibitem{paper3} Sverdlov R 2008 Gauge fields in causal set theory
{\em Preprint\/} 0807.2066

\bibitem{GibSol} Gibbons G W and Solodukhin S 2007 The geometry of small causal diamonds {\em Phys. Lett.} B {\bf649} 317 ({\em Preprint\/} hep-th/0703098)

\bibitem{cc} Sorkin R D 2007 Is the cosmological ``constant" a nonlocal quantum residue of discreteness of the causal set type? {\em AIP Conf. Proc.} {\bf957} 142 ({\em Preprint\/} 0710.1675)

\bibitem{paper5} Sverdlov R 2008 Bosonic fields in causal set theory
{\em Preprint\/} 0807.4709

\bibitem{paper6} Sverdlov R 2008 Spinor fields in causal set theory
{\em Preprint\/} 0808.2956

\bibitem{Rovelli} Rovelli C 2006 Graviton propagator from background-independent quantum gravity {\em Phys. Rev. Lett.} {\bf97} 151301 ({\em Preprint\/} gr-qc/0508124)

\end{thebibliography}
\end{document}